\documentclass[usenatbib]{mnras}
\usepackage{subfig}
\usepackage[T1]{fontenc}
\usepackage{ae,aecompl}
\usepackage{graphicx} 
\usepackage{amsmath}  
\usepackage{amssymb}    
\usepackage{hyperref}

\usepackage{soul}

\makeatletter

\newcommand{\Rmnum}[1]{\expandafter\@slowromancap\romannumeral #1@}
\makeatother
\defcitealias{2016Lieu}{XXL Paper \Rmnum{4}}
\defcitealias{2018Adami}{XXL Paper \Rmnum{20}}
\defcitealias{2016Giles}{XXL Paper \Rmnum{3}}
\defcitealias{2016Lavoie}{XXL Paper \Rmnum{15}}
\defcitealias{2016Pierre}{XXL Paper \Rmnum{1}}
\defcitealias{2016Eckert}{XXL Paper \Rmnum{13}}
\defcitealias{2018Chiappetti}{XXL Paper \Rmnum{27}}
\defcitealias{2016Pacaud}{XXL Paper \Rmnum{2}}
\defcitealias{2016Fotopoulou}{XXL Paper \Rmnum{6}}
\defcitealias{2014Mantz}{XXL Paper \Rmnum{5}}
\defcitealias{2018Mantz}{XXL Paper \Rmnum{17}}
\defcitealias{2018Logan}{XXL Paper \Rmnum{33}}
\defcitealias{2020Ricci}{XXL Paper \Rmnum{44}}

\hypersetup{colorlinks,linkcolor={blue},citecolor={blue}}

\begin{document}
\title[X-ray follow-up of distant XXL clusters]{The XXL Survey: \Rmnum{48}; X-ray follow-up of distant XXL clusters: Masses, scaling relations and AGN contamination\thanks{Based on observations obtained with XMM-Newton, an ESA science mission with instruments and contributions directly funded by ESA Member States and NASA.}}

\author[R. T. Duffy et al.]{
R. T. Duffy,$^{1,2}$\thanks{Email: \texttt{ryan.duffy@cea.fr}}
C. H. A. Logan,$^{1,3}$
B. J. Maughan,$^{1}$
D. Eckert,$^{4}$
N. Clerc,$^{5}$
S. Ettori,$^{6,7}$
\newauthor 
\,\,F. Gastaldello,$^{8}$
E. Koulouridis,$^{9}$
M. Pierre,$^{2}$
M. Ricci,$^{10,11}$
M. Sereno,$^{6,7}$
I. Valtchanov$^{12}$
\newauthor
\,\,and J. P. Willis$^{13}$
\\
$^{1}$HH Wills Physics Laboratory, University of Bristol, Tyndall Avenue, Bristol, BS8 1TL, UK\\
$^{2}$AIM,   CEA,   CNRS,   Universit\'e Paris-Saclay,   Universit\'e  Paris Diderot, Sorbonne Paris Cit\'e, 91191 Gif-sur-Yvette, France\\
$^{3}$European Space Agency (ESA), European Space Astronomy Centre (ESAC), Camino Bajo del Castillo s/n, 28692 Villanueva de la\\ Ca\~{n}ada, Madrid, Spain\\
$^{4}$Department of Astronomy, University of Geneva, Ch. d’Ecogia 16, CH-1290 Versoix\\
$^{5}$IRAP, Universit\'e de Toulouse, CNRS, UPS, CNES, F-31028 Toulouse, France\\
$^{6}$INAF, Osservatorio di Astrofisica e Scienza dello Spazio, via Pietro Gobetti 93/3, 4019 Bologna, Italy\\
$^{7}$INFN, Sezione di Bologna, viale Berti Pichat 6/2, I-40127 Bologna, Italy\\
$^{8}$INAF, IASF-Milano, via A. Corti 12, I-20133 Milano, Italy\\
$^{9}$Institute for Astronomy \& Astrophysics, Space Applications \& Remote Sensing, National Observatory of Athens, GR-15236 Palaia\\ Penteli, Greece\\
$^{10}$Laboratoire d’Annecy de Physique des Particules, Universit\'e Savoie Mont Blanc, CNRS/IN2P3, F-74941 Annecy, France\\
$^{11}$Laboratoire Lagrange, Universit\'e C\^ote d’Azur, Observatoire de la C\^ote d’Azur, CNRS, Blvd de l’Observatoire, CS 34229, 06304\\ Nice cedex 4, France\\
$^{12}$Telespazio Vega UK for ESA, European Space Astronomy Centre, Operations Department, 28691 Villanueva de la Ca\~{n}ada, Spain\\
$^{13}$Department of Physics and Astronomy, University of Victoria, 3800 Finnerty Road, Victoria, V8P 5C2 BC, Canada}

\maketitle
 
\begin{abstract}
We use deep follow-up \emph{XMM-Newton} observations of 6 clusters discovered in the XXL Survey at $z>1$ to gain robust measurements of their X-ray properties and to investigate
the extent to which scaling relations at low redshift are valid at $z>1$. This sample is unique as it has been investigated for AGN contamination, which ensures measurements are not undermined by systematic uncertainties, and pushes to lower mass at higher redshift than is usually possible, for example with Sunyaev-Zel'dovich (SZ) selected clusters. We determine the flux contribution of point sources to the XXL cluster flux in order to test for the presence of AGN in other high-redshift cluster candidates, and find 3XLSS J231626.8-533822 to be a point source misclassified as a cluster and 3XLSS J232737.3-541618 to be a genuine cluster. We present the first attempt to measure the hydrostatic masses in a bright subsample of $z$>1 X-ray selected galaxy clusters with a known selection function. Periods of high particle background significantly reduced the effective exposure times of observations (losing >50\% in some cases) limiting the power of this study. When combined with complementary SZ selected cluster samples at higher masses, the data appear broadly consistent with the self-similar evolution of the low redshift scaling relations between ICM properties and cluster mass, suggesting that properties such as the X-ray temperature, gas mass and SZ signal remain reliable mass proxies even at high redshift.
\end{abstract}

   
\section{Introduction}
Galaxy clusters are the largest gravitationally bound objects within the Universe, and therefore represent the culmination of cosmic structure formation. This makes them powerful tools for studies of cosmology and astrophysics, especially as ideal laboratories to study how large scale structure forms and evolves with time.

Studies of samples of galaxies clusters are typically undertaken at relatively low redshifts. At higher redshifts ($z$>1), X-ray observations have provided many detections of individual clusters \citep{2006Bremer,2011Nastasi,2011Santos}. This includes clusters in the XXL Survey such as XLSSC 122 \citep[][hereafter XXL Paper \Rmnum{17}]{2018Mantz} and XLSSC 102 \citep[][hereafter XXL Paper \Rmnum{44}]{2020Ricci}, however well-defined samples of exclusively high-redshift X-ray-selected clusters are rarer. As a result of this constraints on distant systems, and their evolution to the present, are sparse. At cosmological distances, clusters are hugely outnumbered by active galactic nuclei (AGN). The low surface brightness of the cluster emission and the fact that the angular extent of the emission can be similar to the point spread function (PSF) of some X-ray telescopes (particularly \emph{XMM-Newton}) means that resolving clusters at these distances can be difficult, and often requires follow-up observations with sufficient angular resolution to differentiate them from AGN \citep[][hereafter XXL Paper \Rmnum{33}]{2018Logan}. As the X-ray surface brightness drops rapidly with increasing redshift, and the Sunyaev-Zeldovich (SZ) signal is independent of redshift, population studies at high redshifts tend to rely on X-ray follow-up observations of the clusters detected from wide area sky surveys using the thermal SZ effect \citep{2017Bartalucci,2018Bartalucci,2019Bulbul,2020Lovisari,2020Ghirardini}.

The measurement of galaxy cluster masses is challenging at the best of times, but even more so at $z>1$. Most of the mass of a cluster is contained in dark matter, and thus mass estimation techniques probe the mass of clusters indirectly. In the X-ray regime this is done by looking at the effect of the gravitational potential of the cluster on the intracluster medium (ICM), relying on the assumption of hydrostatic equilibrium. This calculation requires the measurement of both gas density and temperature profiles, and it is impractical to obtain sufficient numbers of photons from high redshift clusters to derive these without lengthy observations. 

Galaxy clusters form via the process of hierarchical structure formation, whereby the largest virialised structures form from primordial density fluctuations amplified through gravitational collapse and mergers of smaller systems. As a consequence of this, galaxy clusters maintain similar properties and appear to be rescaled versions of one another. Assuming a spherically symmetric ICM is heated exclusively by gravitational processes while obeying hydrostatic equilibrium, it is possible to derive scaling relations between the X-ray properties of the ICM. Measuring these properties for samples of clusters and comparing them to self-similarity provides a powerful test for the cosmology of the Universe. It also allows for the derivation of more difficult to obtain parameters such as the cluster mass from more readily obtainable X-ray observables such as the temperature and luminosity. The calibration of X-ray scaling relations with observational data will prove especially important with the advent of eROSITA, where large numbers of clusters will be detected with few photons.

At relatively low redshifts, a number of studies have measured low-scatter scaling relations between various X-ray observables and the cluster mass \citep{2007Arnaud,2009Vikhlinin}. Work on scaling relations featuring clusters at higher redshifts ($z>0.5$), feature only a handful of measured hydrostatic masses for clusters at $z>1$ \citep{2004Ettori,2007Schmidt,2016Amodeo} or use cluster masses derived from the SZ signal \citep{2019Bulbul}. In this work, we present the first attempt to measure the hydrostatic masses in a sub-sample of $z$>1 X-ray selected galaxy clusters with a known selection function.

This work uses the XXL Survey \citep[][XXL Paper \Rmnum{1}]{2016Pierre}, the largest observing programme undertaken by \emph{XMM-Newton}. The survey covers two distinct fields (XXL-North and XXL-South) totalling 50 square degrees. The primary aim is to study the large-scale structure of the Universe using the distribution of clusters of galaxies as tracers for the distribution of matter. To date, the survey has detected hundreds of galaxy clusters out to a redshift of $z\sim2$.

It is especially important to confirm that clusters detected at these redshifts are genuine, and are not significantly contaminated by X-ray emission from unresolved point sources. AGN in galaxy clusters are significantly more common at higher redshift, with at least three times as many detected in clusters at $1<z<1.5$ than in clusters at $0.5<z<1$ \citep{2009Galametz}. A cluster with significant point source contamination will have its flux and temperature overestimated, impacting not only the use of these properties as mass proxies, but also studies of X-ray scaling relations. Measurement of the AGN contamination in high redshift clusters in the XXL Survey, making use of \emph{Chandra} observations to resolve these point sources, has previously been undertaken by \citetalias{2018Logan}.

In this work, we study a bright subsample of clusters with $z>1$, with the primary goal of measuring their hydrostatic masses. We conclude the work of \citetalias{2018Logan} and present the AGN contamination for those $z>1$ clusters which had yet to be observed by \emph{Chandra} at the time of publication of that work. In Section \ref{sec:sample} we introduce the XXL $z>1$ bright cluster sample. In Section \ref{sec:agncontam} we conclude the analysis of the AGN contamination in $z>1$ XXL clusters previously unobserved by \emph{Chandra}. We present the observations used to study the thermodynamic properties and hydrostatic masses of the $z>1$ bright cluster sample and detail the data analysis procedure in Section \ref{sec:dataanalysis}. The results of this work are presented in Section \ref{sec:results}, and finally in Section \ref{sec:conclusions} we give a summary and our conclusions. Throughout this work we assume a $\Lambda$CDM cosmological model, with Hubble parameter $H_{0}$ = 70 km s$^{-1}$ Mpc$^{-1}$, matter density $\Omega_{\rm m}$ = 0.3 and dark energy density $\Omega_{\Lambda}$ = 0.7.

\subsection{XXL $z>1$ Bright Cluster Sample}
\label{sec:sample}
XXL is a survey undertaken by \emph{XMM-Newton} split between two fields and optimised for the discovery of galaxy clusters. Of the cluster candidates detected so far \citep[][hereafter XXL Paper \Rmnum{20}]{2018Adami}, 17 are found to be at $z$ > 1. After the application of a flux cut of $10^{-14}$ erg s$^{-1}$ cm$^{-2}$ in the 0.5-2.0 keV band, we are left with a bright subsample of 7 cluster candidates, referred to as the $z > 1$ bright sample. Prior to this work, 6 of the cluster candidates in the subsample were observed with \emph{Chandra}. The higher spatial resolution of \emph{Chandra} was used to characterise the AGN contamination fraction in each cluster candidate. In each of these cases, the cluster candidates were found to be free from significant AGN contamination, and confirmed as clusters \citepalias{2018Logan}. These were then targeted for longer observations with \emph{XMM-Newton} where necessary, with the final cluster candidate (3XLSS J231626.8-533822) to follow once its \emph{Chandra} observation had been analysed. We discuss the AGN contamination of this cluster, along with 3XLSS J232737.3-541618 (which is not included in the $z>1$ bright sample) in the following sections.

\section{AGN Contamination of $z>1$ XXL Clusters}
\label{sec:agncontam}
\subsection{Background and Dataset}
At the time of the publication of \citetalias{2018Logan}, there were three cluster candidates at $z>1$ which were unobserved by \emph{Chandra} and thus not included in the paper. Since then, the redshift of one cluster candidate (3XLSS J233116.6-550737) has been revised to $z=0.61$. The redshift was updated as a result of Gemini multi-object spectroscopy and photometric redshifts of candidate cluster galaxies, which showed two overlapping structures within the X-ray contours: one at $z\sim0.61$ and one at $z\sim1.3$. Consequently, 3XLSS J233116.6-550737 is no longer part of the $z>1$ XXL cluster sample. The sample information for the two remaining cluster candidates is presented in Table \ref{table:chandra_snapshots_sample_summary}\footnote{The tables in this section are very similar to those in \citetalias{2018Logan}, however, the column giving the cluster class from \citet{2013Willis} is now excluded. The two cluster candidates covered here are in the XXL-S field, and are consequently not in the XMM-LSS field studied by \citet{2013Willis}.}.

\begin{table*}
    \begin{center}
    \scalebox{0.85}{
    \begin{tabular}{lcccccccccc}
        \hline
        XXLID & ObsID & Class & $z$ & RA & Dec. &  F$_{60}$  & Chip  & Clean time\\
            &  & XXL & & (J2000) & (J2000) & ($10^{-14}$ erg/cm$^{2}$/s)  & configuration & (ksec) \\
        \hline
        3XLSS J231626.8-533822 & 20537 & C2 & 1.28 & 349.111 & -53.639 & 2.0$\pm$0.4 & ACIS-S & 9.9 \\
        3XLSS J232737.3-541618 & 20533 & C2 & 1.02 & 351.906 & -54.272 & 1.0$\pm$0.3 & ACIS-S & 9.9 \\
        \hline
    \end{tabular}
    }
    \caption{Summary of the cluster candidate sample and \emph{Chandra} data. Column 1 is the cluster candidate name; column 2 is the \emph{Chandra} ObsID; column 3 is the cluster class from the XXL pipeline consistent with the classes given in \citetalias{2018Logan} and \citetalias{2018Adami} (C1 clusters are expected to be mostly free of contamination by point sources, while the C2 sample is expected to contain 50\% misclassified AGN);   
    column 4 is the redshift of the cluster candidate (we note that both cluster candidates have photometric redshifts; in addition, 3XLSS J231626.8-533822 has galaxies at z=1.28, but spectroscopic confirmation is pending, since we have spectroscopic redshifts for only two galaxies within 500 kpc of the X-ray peak, whereas spectroscopic redshifts of three galaxies are required by XXL for spectroscopic confirmation); columns 5 and 6 are the RA and Dec. coordinates of the cluster centre \citepalias{2018Adami}; column 7 is the cluster flux in the 0.5 - 2 keV energy band measured in the 60$''$ cluster region using XXL data (neither cluster candidate is included in \citetalias{2018Adami} and so their cluster fluxes were computed directly using a growth curve analysis, following the method described in \citealt[XXL Paper \Rmnum{2}]{2016Pacaud}); column 8 is the CCD chip configuration for the observation; column 9 is the cleaned \emph{Chandra} observation time. We note that both cluster observations were targeted, and as such they were observed on-axis by \emph{Chandra}. 
    }
    \label{table:chandra_snapshots_sample_summary}
\end{center}
\end{table*}

\subsection{Analysis}
We closely follow the prescription of \citetalias{2018Logan}, to which the reader should refer for more detail. The analysis methods used in that paper are briefly outlined in this section. 

The two cluster candidates were analysed with the \textsc{CIAO}\footnote{\url{https://cxc.cfa.harvard.edu/ciao/}} 4.9 software package and \textsc{CALDB}\footnote{\url{https://cxc.cfa.harvard.edu/caldb/}} version 4.7.4 \citep{2006Fruscione}, consistent with the packages used in \citetalias{2018Logan}. The level 1 event files were reprocessed using the \textsc{CIAO} \texttt{chandra\_repro} tool. We then identified (and subsequently removed) periods of background flaring using lightcurves analysed with the \texttt{deflare} tool.

In order to detect point sources in the \textit{Chandra} observations, we used the \textsc{CIAO} \texttt{wavdetect} tool on images in the 0.3 - 8 keV band, and then used the \textsc{CIAO} \texttt{srcflux} tool to estimate fluxes of any detected point sources (we note that the point source fluxes, as presented in Table \ref{table:chandra_snapshots_results_summary} and \ref{table:chandra_snapshots_indivptsrcs_summary} are measured in the 0.5 - 2 keV band). When measuring the fluxes, the source region used was the 90\% encircled energy radius of the PSF at 1 keV, and the background region was an annulus (centred on the same coordinates as the source region) with the inner radius equal to the source radius, and the outer radius set to five times the inner radius. To model the point source flux we assumed a power law model with $\Gamma$ = 1.7, consistent with the modelling used in \citet[][XXL Paper \Rmnum{6}]{2016Fotopoulou}. We also checked for any other potential point sources not detected by \texttt{wavdetect} by searching for any points with (i) at least 4 counts in a single pixel, or (ii) at least 6 counts in a 1\arcsec\, circle with at least one pixel containing 2 or more counts, although neither of the two cluster candidates in this paper had any additional point sources detected using this alternative method. Optical images were also used to check for any optical counterparts to the point sources detected in the X-ray images.

The \texttt{srcflux} tool was also used to constrain the flux of any extended emission from cluster candidates in the 0.5 - 2 keV band, with a 60\arcsec\, radius circle used as the source region, and a 120-180\arcsec\, annulus used as the background region. We assumed an absorbed APEC thermal plasma model \citep{2001Smith} to model the flux, and set the metal abundance set to 0.3 solar, and the plasma temperature to 3.5 keV.

\subsection{Results}
\begin{figure*}
	\centering
	\begin{tabular}{cc}
	\subfloat[3XLSS J231626.8-533822 Optical]{\includegraphics[height=1.5in]{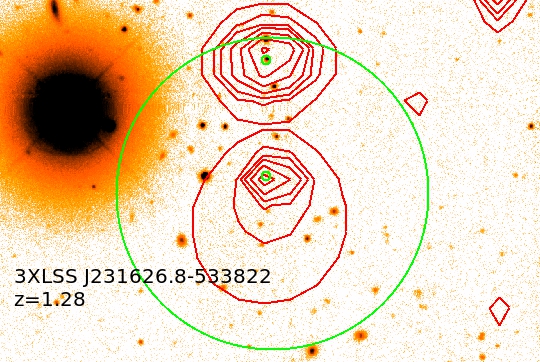}
	} &
	\subfloat[3XLSS J231626.8-533822 \textit{Chandra} Smoothed]{\includegraphics[height=1.5in]{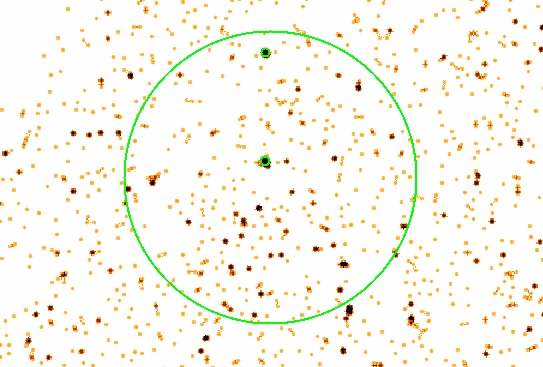}} \\
	\subfloat[3XLSS J232737.3-541618 Optical]{\includegraphics[height=1.5in]{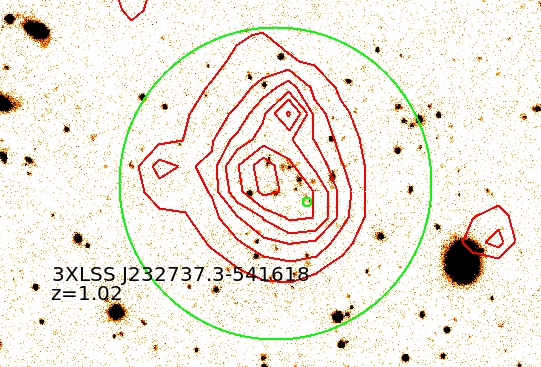}
	} &
	\subfloat[3XLSS J232737.3-541618 \textit{Chandra} Smoothed]{\includegraphics[height=1.5in]{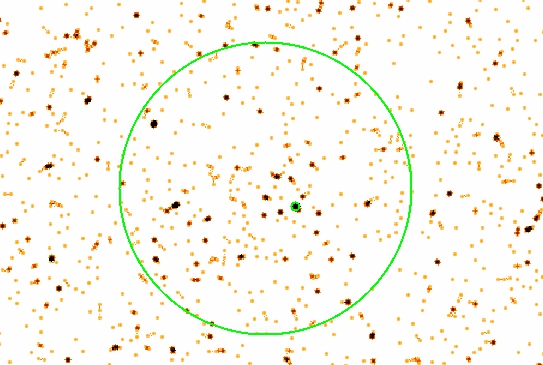}} \\
	\\
	\end{tabular}
	\caption{Images in the top row correspond to 3XLSS J231626.8-533822, while images in the bottom row correspond to 3XLSS J232737.3-541618. The left column shows 0.5-2.0 keV band \emph{XMM-Newton} contours (red) overlaid on optical i-band images from CFHTLS for each of the clusters. The right column shows 0.3-8.0 keV \emph{Chandra} images smoothed using a Gaussian with $\sigma \sim 2.5\arcsec$. The large green circle in each image corresponds to a 60$\arcsec$ circle centred on the cluster centre. Point sources within 60$\arcsec$ of the centre are indicated with the smaller green circles. For 3XLSS J231626.8-533822, the northernmost point source was previously detected by \emph{XMM-Newton}. In each image north is up and east is to the left.}
	\label{fig:chandra_snapshot_images}
\end{figure*}

Our results are presented mimicking the format of the tables in \citetalias{2018Logan} are shown in Table \ref{table:chandra_snapshots_results_summary} and \ref{table:chandra_snapshots_indivptsrcs_summary}. Images of the clusters and further results are shown in Figure \ref{fig:chandra_snapshot_images}.

\begin{table*}
    \begin{center}
    \scalebox{0.70}{
        \begin{tabular}{lcccccccccc}
            \hline
            XXLID & Class & $z$ &  F$_{60}$  & No. of point &  \emph{Chandra} point source flux  & AGN contamination & Final & \emph{Chandra} cluster flux\\
            &  XXL    & &  ($10^{-14}$ erg/cm$^{2}$/s) & sources & ($10^{-14}$ erg/cm$^{2}$/s) & fraction & assessment & ($10^{-14}$ erg/cm$^{2}$/s)\\
            \hline
            3XLSS J231626.8-533822 & C2 & 1.28 & 2.2$\pm$0.5 & 1 & 1.94$\pm$0.42 & 0.88 & FC & $<$0.80 \\
            3XLSS J232737.3-541618 & C2 & 1.02 & 0.8$\pm$0.4 & 1 & 0.17$\pm$0.13 & 0.21 & PC & $2.55^{+0.99}_{-0.94}$ \\
            \hline
        \end{tabular}
        }
    \caption{Summary of point source detection and cluster contamination from the \textit{Chandra} data. The \emph{Chandra} cluster flux measurement is also shown. Column 4 is the XXL cluster flux. Column 5 gives the number of point sources detected by \texttt{wavdetect} within a 60$''$ radius region around the cluster centre that were not  previously detected by XXL
. Column 6 gives the total flux of all of the point sources detected by \texttt{wavdetect} within a 60$''$ region around the cluster centre that weren't detected by XXL,
with the 1$\sigma$ lower and upper limits are given as error. All fluxes are in the 0.5 - 2 keV energy band. Column 7 gives the fraction of F$_{60}$ resolved into point sources by \textit{Chandra}, as described in \citetalias{2018Logan} Section 3.1. Column 8 gives our assessment of the cluster. Column 9 is the cluster flux as calculated from \emph{Chandra} data after point source removal (described in \citetalias{2018Logan} Section 3.2) with 1$\sigma$ errors. Individual point source fluxes and positions are given in Table \ref{table:chandra_snapshots_indivptsrcs_summary}.}
    \label{table:chandra_snapshots_results_summary}
    \end{center}
\end{table*}

\begin{table*}
    \begin{center}
    \scalebox{0.88}{
        \begin{tabular}{lcccccccccc}
            \hline
            XXLID   & Class & $z$ & RA & Dec. & Flux   & Resolved  & Separation from    \\
            &   XXL   & & (J2000) & (J2000) & ($10^{-14}$ erg/cm$^{2}$/s) & by \emph{XMM}  & cluster centre ($''$)  \\
            \hline
            3XLSS J231626.8-533822 & C2 & 1.28 & 349.112 & -53.637 & 1.94$^{+0.44}_{-0.39}$ & No & 7 \\
            & & & 349.112 & -53.625 & 1.23$^{+0.36}_{-0.30}$ & 3XLSS J231626.8-533728 & 51 \\
            3XLSS J232737.3-541618 & C2 & 1.02 & 351.900 & -54.274 & 0.17$^{+0.10}_{-0.16}$ & No & 14 \\
            \hline
        \end{tabular}
    }
    \caption{Summary of the fluxes for all point sources within 60\arcsec\, of the cluster centre
    . Column 6 is the individual point source flux as calculated from the \emph{Chandra} data with 1$\sigma$ errors. All fluxes are in the 0.5 - 2 keV energy band. 
    Column 7 states whether the \emph{Chandra} detected point source was previously resolved by XXL and thus excluded from the F$_{60}$ measurements; for cases where the point source was resolved by XXL, its name as in \citet[][XXL Paper \Rmnum{27}]{2018Chiappetti} is provided.}
    \label{table:chandra_snapshots_indivptsrcs_summary}
\end{center}
\end{table*}

Results for each of the two clusters are summarised below:
\begin{itemize}
    \item 3XLSS J231626.8-533822 / 20537 / C2 / FC (fully contaminated): \newline This cluster candidate has two point sources detected in our \textit{Chandra} data in the 60\arcsec\, cluster region. One of these point sources was not previously detected by \textit{XMM-Newton}, and accounts for 88\% of the previous XXL cluster flux estimate. This point source also has an optical counterpart, clearly visible in Figure \ref{fig:chandra_snapshot_images}. The previously detected point source also has an optical counterpart, though it is less bright in the optical image. We also attempted to constrain the flux of any extended emission using the \textit{Chandra} data (see Table \ref{table:chandra_snapshots_results_summary}), masking the additional \textit{Chandra} point source (as well as the original point source found by \textit{XMM-Newton}), and compute only a (3 $\sigma$) upper limit to the cluster flux of $0.8 \times 10^{-14}$ erg/cm$^{2}$/s. This further suggests that the original XXL cluster flux originated completely from this previously undetected point source. Thus, we conclude that this cluster candidate is fully contaminated.
    \newline
    \item 3XLSS J232737.3-541618 / 20533 / C2 / PC (partially contaminated): \newline This cluster candidate has one point source detected in our \textit{Chandra} data in the 60\arcsec\, cluster region. This point source was not previously detected by \textit{XMM-Newton}, but only accounts for 21\% of the previous XXL cluster flux estimate. This point source also has an optical counterpart. We also measured the cluster flux using the \textit{Chandra} data (see Table \ref{table:chandra_snapshots_results_summary}), masking the additional \textit{Chandra} point source (as well as the original point source found by \textit{XMM-Newton}), and compute the cluster flux to be $2.55^{+0.99}_{-0.94} \times 10^{-14}$ erg/cm$^{2}$/s. The small level of contamination of the original XXL cluster flux from the point source detected from the \textit{Chandra} data, along with the measured \textit{Chandra} cluster flux, strongly suggests genuine emission originating from the ICM. Thus, we conclude that this cluster is a partially contaminated cluster. We note that as the \textit{XMM-Newton} observation and \textit{Chandra} observations were taken years apart, there is a possibility of variability between observations \citep{2019Maughan}. Over this timescale, a variation of $\sim$50\% is possible, though even if the AGN has decreased in brightness since the \textit{XMM-Newton} observation, the AGN would not dominate, and our conclusion of a partially contaminated cluster would still be correct.
\end{itemize}

As a result of this analysis, we have shown that 3XLSS J231626.8-533822 is not a cluster. Thus, for the rest of this work, we are utilising a subsample of 6 clusters. In the next sections we move on to presenting the measurement of thermodynamic properties and hydrostatic masses in this subsample of X-ray selected clusters at $z>1$.

\section{\emph{XMM-Newton} Analysis of the $z>1$ Bright Cluster Sample}
\label{sec:dataanalysis}

\subsection{Processing}
The $z>1$ bright clusters and their relevant \emph{XMM-Newton} exposures are given in Table \ref{table:obs}.

\begin{table*}
    \centering
	\begin{tabular}{l c c c c c c c c}
	    \hline
		Cluster & ObsID & $z$ & Flux & pn $t_{\rm exp}$ & MOS1 $t_{\rm exp}$ & MOS2 $t_{\rm exp}$ \\
		& & & ($10^{-14}$ erg/cm$^{2}$/s) & (ks) & (ks) & (ks) \\
		\hline
        XLSSC 029 & 0210490101 & 1.05 & $3.2\pm0.2$ & 66.4 & 82.6 & 83.1 \\ 
		XLSSC 048 & 0821250601 & 1.01 & $1.1\pm0.2$ & 19.5 & 30.2 & 31.6 \\ 
		 & 0821250701 & & & 29.5 & 47.7 & 51.3 \\
		XLSSC 072 & 0673110201 & 1.00 & $4.1\pm0.4$ & 23.7 & 31.4 & 31.4 \\ 
		XLSSC 122 & 0760540101 & 1.98 & $1.3\pm0.2$ & 79.0 & 90.6 & 90.9 \\ 
		XLSSC 634 & 0821250401 & 1.08 & $4.8\pm0.5$ & 14.6 & 24.5 & 24.3  \\ 
		3XLSS J021325.0-042000 & 0821250501 & 1.20 & $1.8\pm0.3$ & 35.2 & 54.0 & 59.6 \\ 
		\hline
	\end{tabular}
	\caption{Clusters and \emph{XMM-Newton} observations used in this work. Column 1 gives the cluster name; Column 2 gives the \emph{XMM-Newton} observation ID; Column 3 gives the cluster redshift; Column 4 gives the X-ray flux within 60$\arcsec$ of the cluster centre in the 0.5-2.0 keV band and Columns 5-7 give the clean exposure time for each EPIC detector.}
	\label{table:obs}
\end{table*}

Observations were analysed using SAS version 16.1.0 and the Current Calibration Files (CCF) dated June 2019. Filtered event files are generated for each of the cameras using the ESAS\footnote{\url{https://heasarc.gsfc.nasa.gov/docs/xmm/esas/cookbook/xmm-esas.html}} tasks \texttt{mos-filter} and \texttt{pn-filter}. \texttt{mos-filter} and \texttt{pn-filter} create light curves and a high-energy count rate histogram from the observation's field-of-view data. They then fit a Gaussian to the peak count rate and determine thresholds at $\pm 1.5 \sigma$, creating good time intervals (GTI) files containing time intervals within the thresholds.

Much of observations 0821250501, 0821250601 and 0821250701 were heavily effected by flaring. This is especially true for 0821250501 and 0821250601 where upwards of 50\% of the total exposure was affected on some detectors. After light curve cleaning, there remains some residual soft proton contamination in the field-of-view. To account for this, we measure the soft proton contamination for each observation by calculating the fraction of the counts which are contaminated in the exposed and unexposed portion of the detector between 6 keV and 12 keV for the MOS cameras, and between between 5 keV and 7.3 keV and 10-14 keV for the pn \citep{2008Leccardi}. If the ratio of contaminated counts in the field of view to out of the field of view exceeds 1.15, additional components are added to the background model during spectral fitting.

\subsection{Imaging}
\begin{figure*}
	\centering
	\begin{tabular}{cc}
	\subfloat[XLSSC 029]{\includegraphics[height=2.0in]{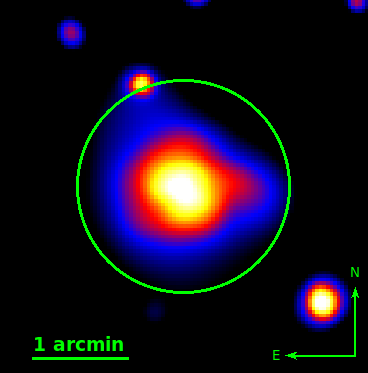}} &
	\subfloat[XLSSC 048]{\includegraphics[height=2.0in]{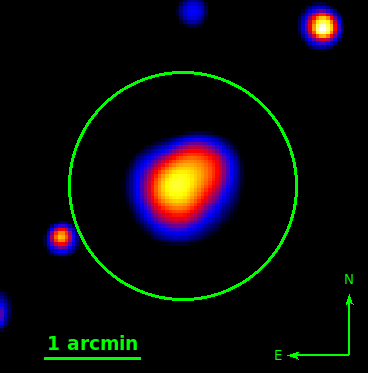}} \\
	\subfloat[XLSSC 072]{\includegraphics[height=2.0in]{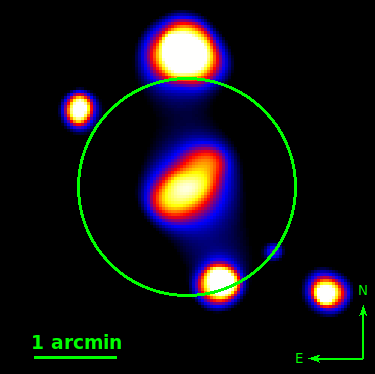}} &
	\subfloat[XLSSC 122]{\includegraphics[height=2.0in]{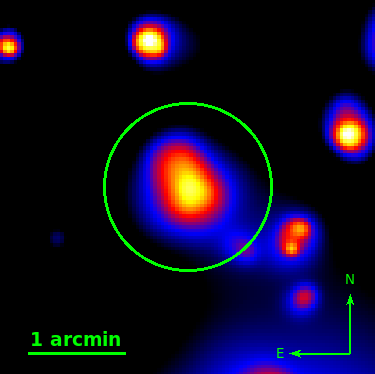}} \\
	\subfloat[XLSSC 634]{\includegraphics[height=2.0in]{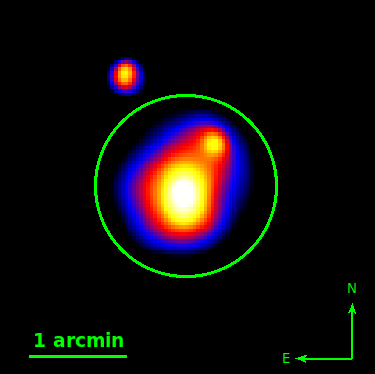}} &
	\subfloat[3XLSS J021325.0-042000]{\includegraphics[height=2.0in]{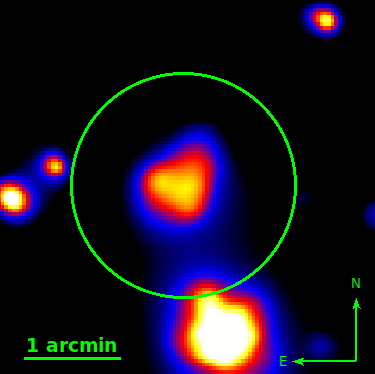}} \\
	\\
	\end{tabular}
	\caption{Adaptively smoothed \emph{XMM-Newton} 0.5-2.0 keV images of each cluster in the subsample. The relevant cluster is in the centre of each frame, inside a circle with radius $r_{500}$.}
	\label{fig:clusterimg}
\end{figure*}

Images for each of the three EPIC detectors (MOS1, MOS2 and pn) are generated from their respective filtered event files in the soft band (0.5 keV - 2 keV). A combined image of each observation is then made by summing the images from each individual camera. The task \texttt{eexpmap} is used to compute exposure maps while also taking the vignetting effect into account. To detect unwanted point and extended sources, we use the XMM-SAS tool \texttt{edetect\_chain} on each of the individual images. This outputs a list of detected sources, which is then converted into a region file. The sources are excised from all subsequent analysis, although remain visible in the adaptively smoothed images shown in Figure \ref{fig:clusterimg}. To produce the adaptively smoothed images we use \textsc{CIAO} 4.10 and the task \texttt{csmooth}. A sub-image of the detector plane centred on the cluster from the combined image is smoothed with a minimum signal-to-noise of 3. The resulting scale map from this process is then used to smooth the same region of a combined exposure map, which is then used to exposure correct the adaptively smoothed image.

\subsection{Spectral Analysis}
In performing the spectral analysis of the clusters, we follow the prescription of \citet[][XXL Paper \Rmnum{3}]{2016Giles}, but the details are summarised here. To account for the background components in the spectral analysis, we first fit a spectral model to a local background. As each of these observations are targeted observations, the local background is taken from an annulus centred on the aiming point of the observation, with a width equal to the extent of the cluster emission. Point sources are excluded from all spectral fits. For our purposes, the extent of the cluster emission was defined as the radius beyond which no significant cluster emission is detected using a threshold of 0.5$\sigma$. The low significance is chosen to ensure that no faint cluster emission is included in the background region. Spectra were then extracted using \texttt{mos-spectra} and \texttt{pn-spectra} and fitted with models to fit the cosmic X-ray background (represented by an APEC plus an additional absorbed APEC + power-law component), the solar wind charge exchange (represented by Gaussians centred on the appropriate energy) and included an additional broken power-law component if the soft proton contamination was high. Parameters were fitted in XSPEC version 12.10.0c across all three cameras simultaneously for clusters with only single observations, and across every camera used in all observations where multiple cameras are used. The spectra were binned to have at least 5 counts per bin and were consequently fitted using the cstat statistic. 

On-source fits were composed of the background model plus an additional absorbed APEC model component \citep{2001Smith}, with the absorbing column fixed to the Galactic value obtained by summing the atomic gas density and molecular hydrogen column density (NHTOT) \citep{2013Willingale} and the abundance table from \citet{2009Asplund}. In order to remain consistent with other papers based on the XXL sample, cluster temperatures we provide temperatures derived within 300 kpc for each cluster, denoted as $T_{300kpc}$. Temperatures are also measured within $r_{500}$ for each cluster, to enable comparisons with high redshift clusters and scaling relations from different work. Throughout the spectral analysis the abundance of the cluster ICM is fixed at 0.3$Z_{\odot}$.

\subsection{Mass Calculation}
To calculate cluster masses we use the backward fitting method \citep{2010Ettori} as presented in \citet{2019Ettori}. This method assumes a parametric mass model with few free parameters, and minimises a likelihood function by comparing predicted and observed temperature profiles in order to constrain the free parameters. We assume that for each cluster the ICM is in hydrostatic equilibrium and that the cluster is fully virialised.

\subsubsection{Density and Temperature Profiles}
To calculate density profiles we first generate surface brightness profiles. These are measured from annular bins with at least 15 total counts in the 0.5-2.0 keV band, which have radius at least 5\% larger than the previous bin. Background counts estimated from the spectral modelling of the background are subtracted from the total counts, and the profiles are corrected for vignetting by dividing by the corresponding exposure in each annulus. The electron density for each cluster is then recovered from the deprojection of the surface brightness profile as described in \citet{2020Eckert}, although we summarise the method here. The observed surface brightness profile is fit to a function which is a linear combination of a large number of $\beta$-profiles, each of which can be individually deprojected. This model is convolved with the PSF, and the best fitting surface brightness profile is found by maximising a likelihood function \citep[their Equation 7]{2020Eckert} using the Python package \texttt{PyMC3} \citep{2016Salvatier}. To convolve the model with the PSF, we create a mixing matrix by establishing the amount of emission from each annulus contributing to other annuli in the profile.


The temperature profiles are generated by creating annular bins with a minimum of 300 net counts in the 0.5-2.0 keV band, moving outwards from the cluster centre. Bins must have a minimum signal-to-noise ratio of 10 to be included in spectral fitting. Fitting the spectra obtained from these regions gives a projected temperature profile for each cluster. When the temperature is low, it is easier to constrain with fewer counts due to additional information in the spectrum from the emission lines, explaining the smaller errors in the outer bins. Due to the substantial flaring in some observations, the maximum number of bins found in for a temperature profile in this work is 4, for clusters XLSSC 029 and 072. XLSSC 634 has 3 bins in its temperature profile, while XLSSC 048, XLSSC 122 and 3XLSS J021325.0-042000 have 2 bins.

\subsection{Fitting and Priors}
Half the sample has only two bins in their temperature profiles, so we use the backward fitting method. The mass model is described by few parameters, and provides a physically motivated extrapolation. To model the mass, we assume an NFW profile \citep{1997Navarro}, a well tested and widely used model for mass profiles of galaxy clusters \citep{2013Ludlow,2018Bartalucci}. It is described by just two parameters, $r_{s}$, a characteristic radius, and $c$, the concentration parameter and is defined as:
\begin{align}
M(<r) = \frac{4}{3}\pi \Delta \rho_{c} r_{s}^{3} f_{c} F(x)
\label{eqn:nfw}
\end{align}

\noindent where

\begin{align*}
f_{c} = \frac{c^{3}}{{\rm ln}(1+c)-c/(1+c)}
\end{align*}

\noindent and for an NFW

\begin{align*}
F(x) = {\rm ln}(1+x)-\frac{x}{1+x}
\end{align*}

\noindent where $x=r/r_{s}$ and $\rho_{c}$ is the critical density at the cluster's redshift and $\Delta$ is the selected overdensity, chosen such that $r_{\Delta}=r_{s}c$. Throughout this work, we use $\Delta=500$.

$r_{s}$ and $c$ are constrained by minimising a likelihood function which compares predicted and observed temperature profiles. A predicted temperature profile is calculated from the inversion of the hydrostatic mass equation, using the gas density profile and the NFW profile:
\begin{align}
P_e(r) = n_e(r) kT(r) = P_0 + \int_{r}^{r_0} \frac{G M_{tot,model}(<r)}{r^2} ~dr
\end{align}
\noindent where $n_{e}(r)$ is the electron density profile, $kT(r)$ is the temperature profile and $P_0$ is an integration constant which represents the pressure at the outer boundary of the cluster. The resultant predicted temperature profiles are marginalised over the value of the $P_0$.

The predicted profile is then projected using the methods described in \citet{2004Mazzotta}, and this predicted profile is then fit to the observed temperature profile. For the fitting process we use a Markov Chain Monte Carlo (MCMC) approach based on the tool \texttt{emcee} \citep{2013Foreman}. In order to break the degeneracy between $c$ and $r_{s}$, we fit for $r_{500}$ and $c_{500}$, with $r_{500}=c_{500}r_{s}$. We use normal priors for each parameter, with $r_{500}\sim \mathcal{N}(500,500)$ truncated at 0 and $c_{500}\sim \mathcal{N}(3,2)$.

\begin{figure*}
	\centering
	\begin{tabular}{cc}
	\subfloat{\includegraphics[height=2.0in]{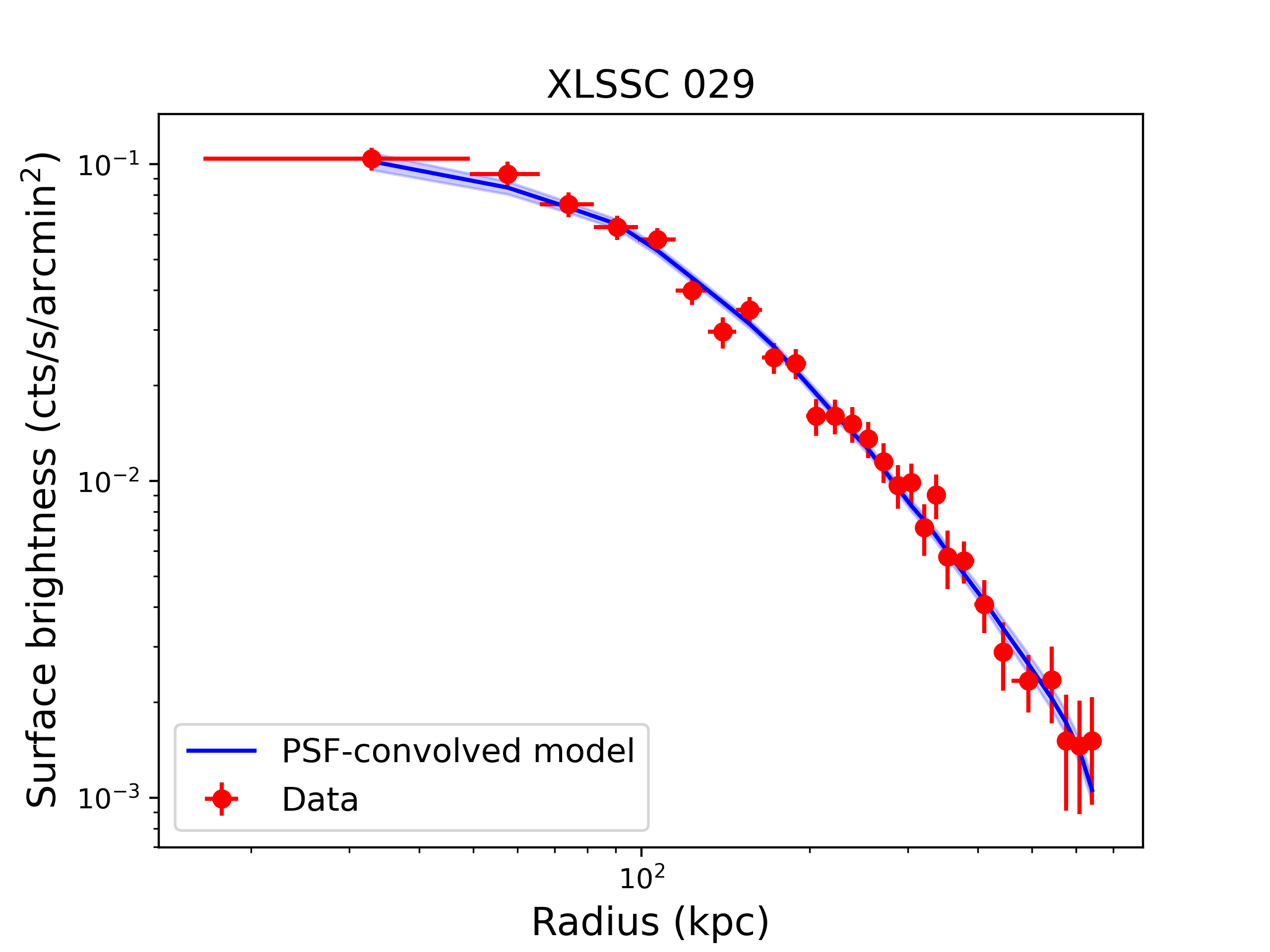}} &
	\subfloat{\includegraphics[height=2.0in]{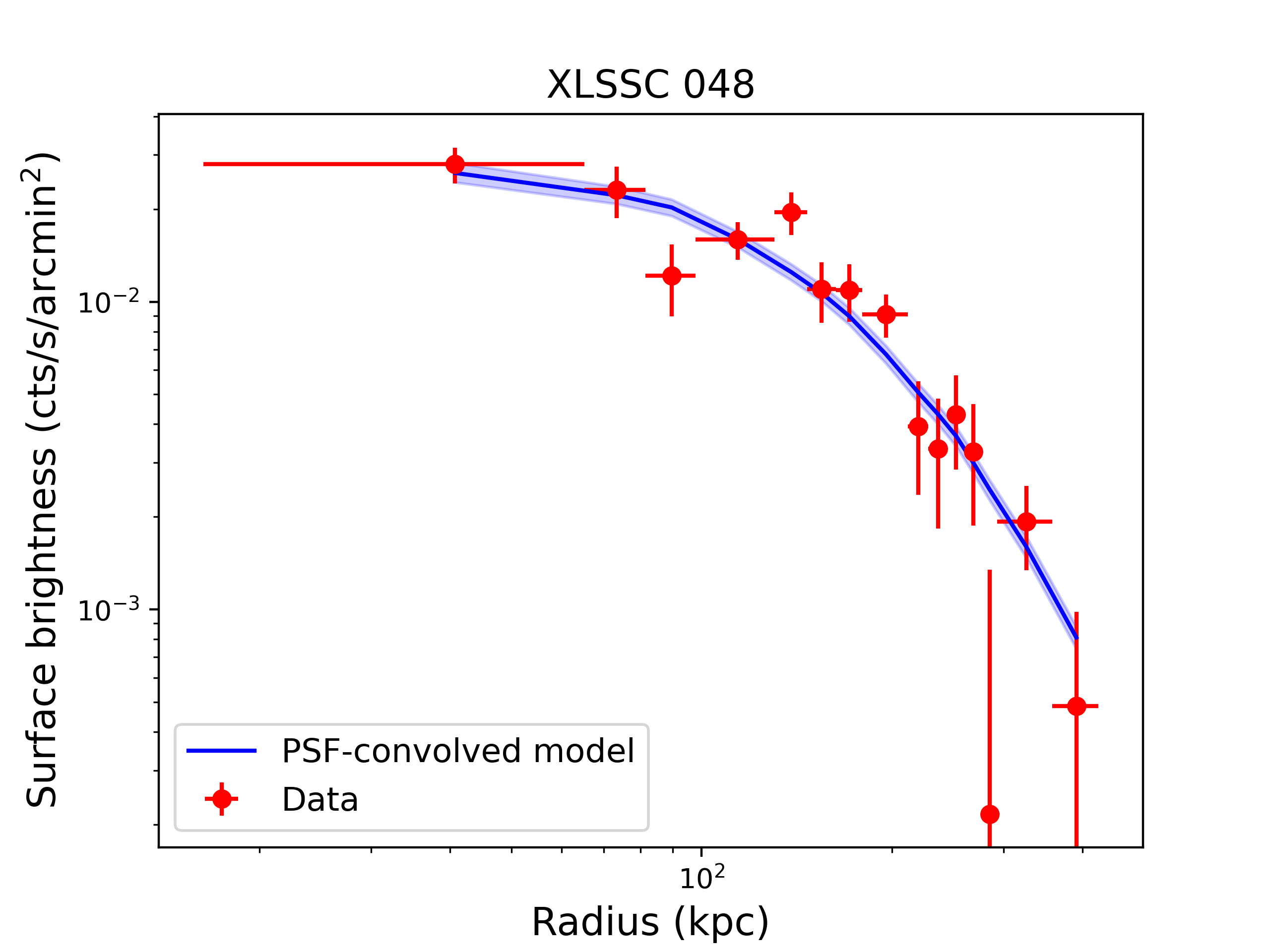}} \\
	\subfloat{\includegraphics[height=2.0in]{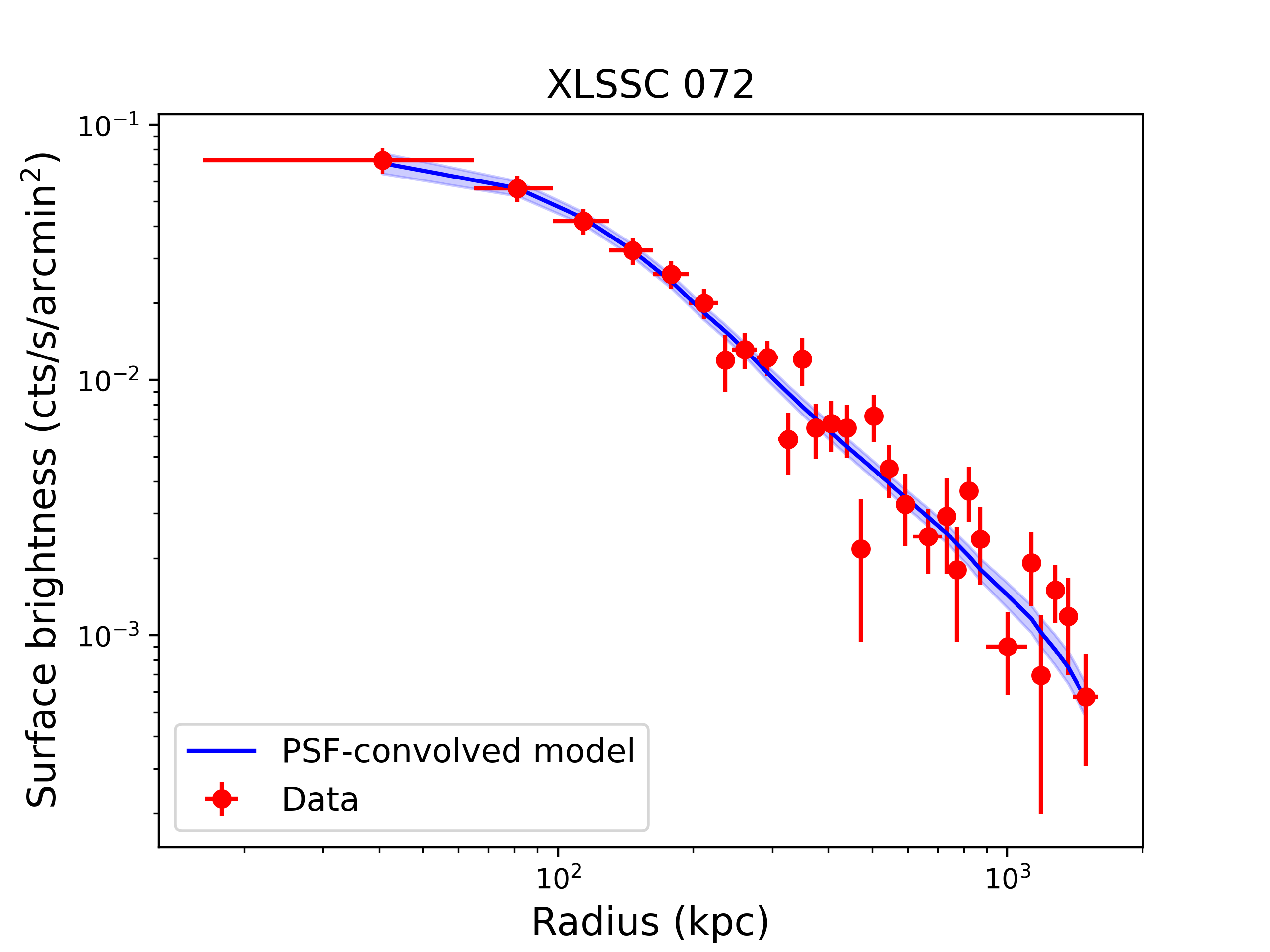}} &
	\subfloat{\includegraphics[height=2.0in]{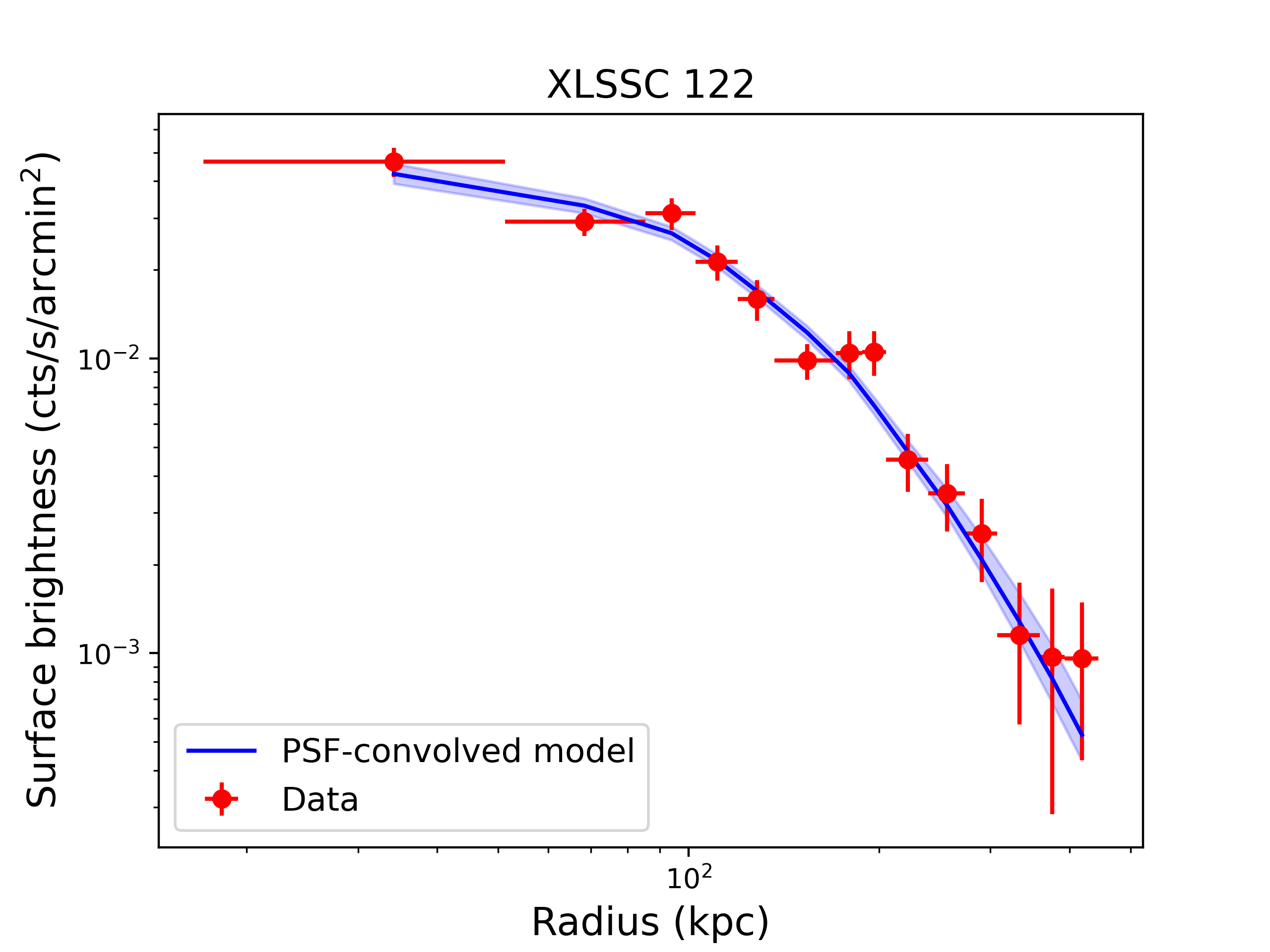}} \\
	\subfloat{\includegraphics[height=2.0in]{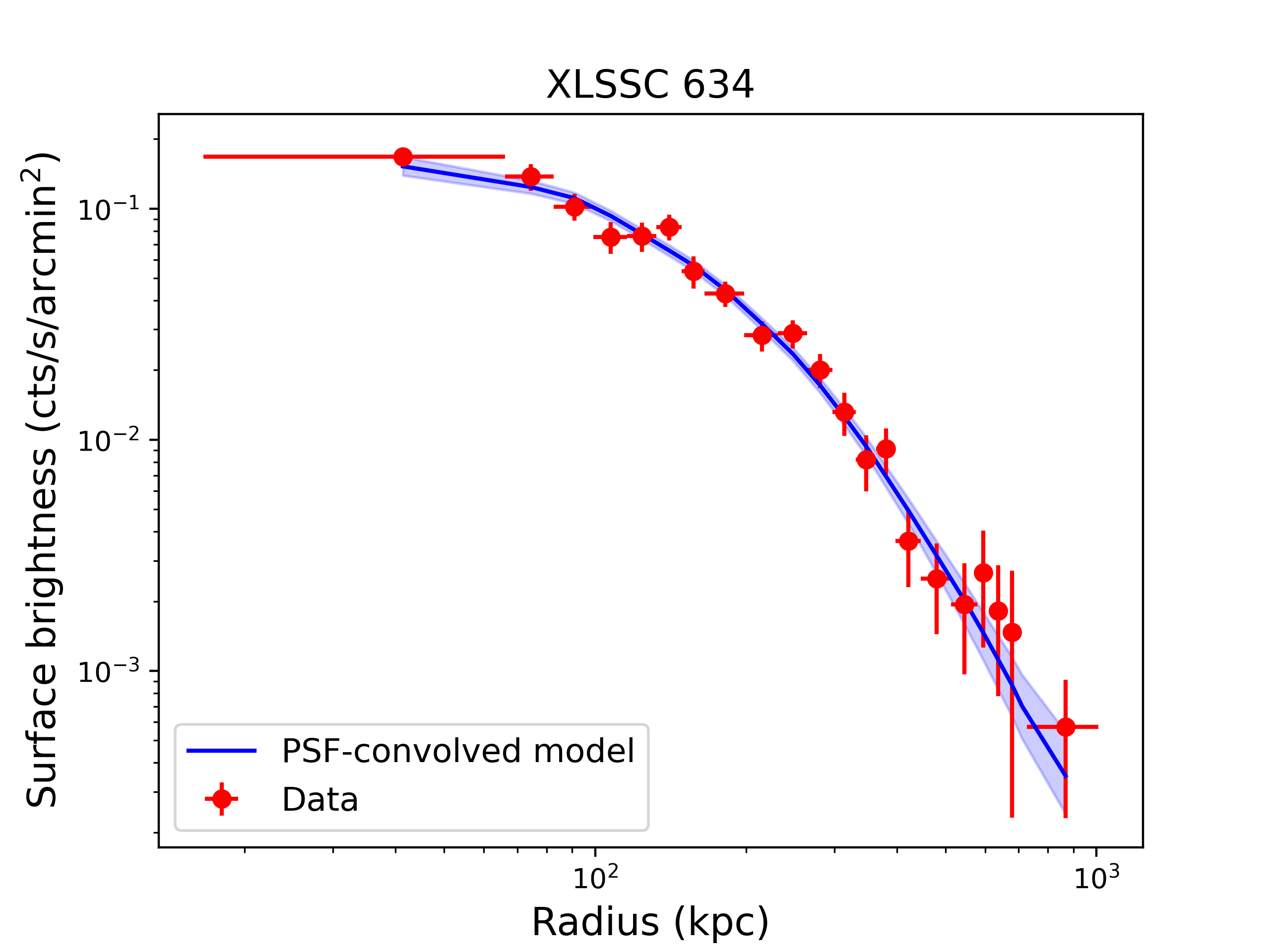}} &
	\subfloat{\includegraphics[height=2.0in]{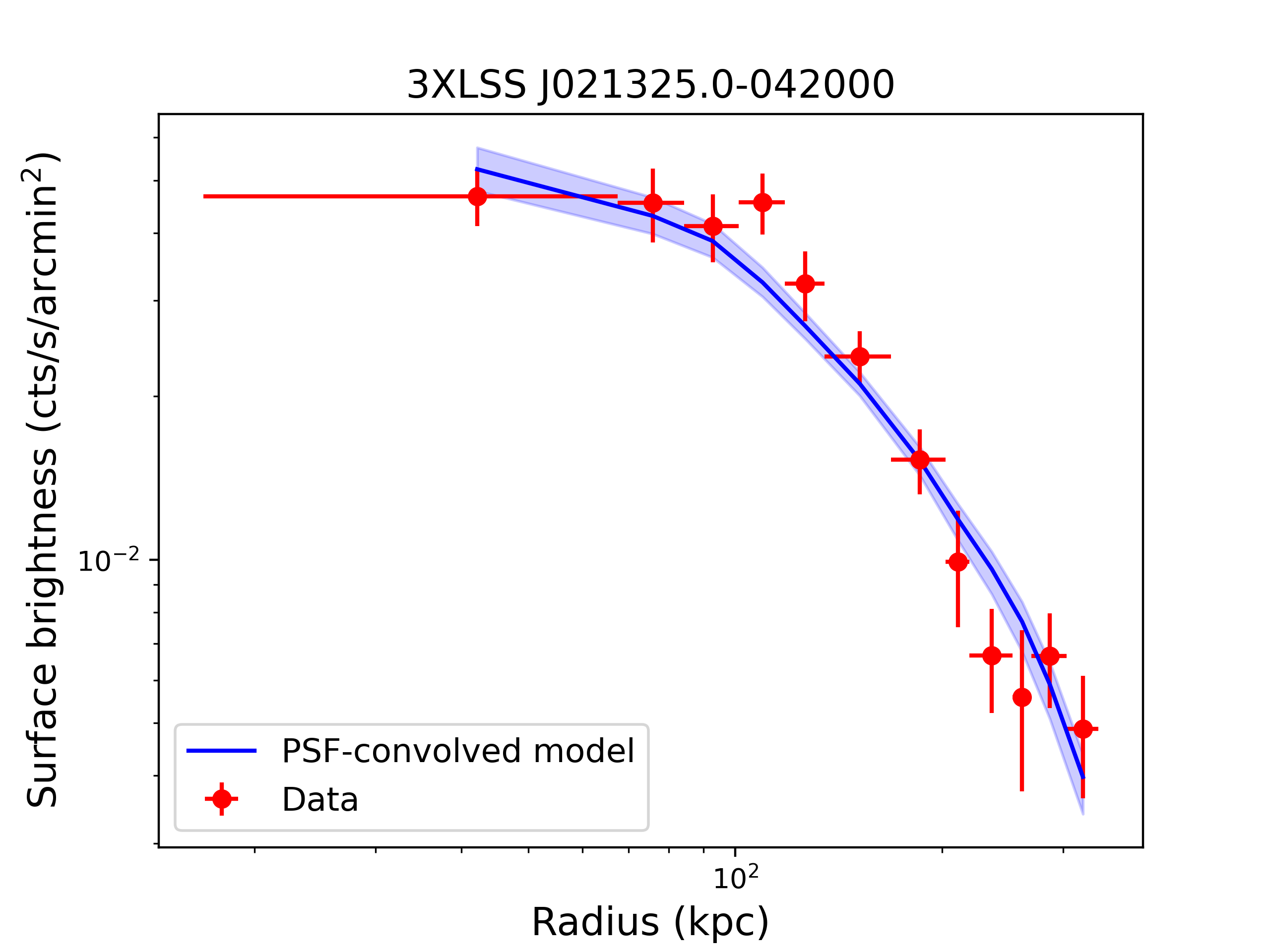}} \\
	\\
	\end{tabular}
	\caption{Surface brightness profiles in the 0.5-2.0 keV band for each of the clusters in the $z>1$ bright sample. The red points indicate the observed surface brightness profile, the blue lines represent the PSF-convolved model.}
	\label{fig:EMprofs}
\end{figure*}

\begin{figure*}
	\centering
	\begin{tabular}{cc}
	\subfloat{\includegraphics[height=2.0in]{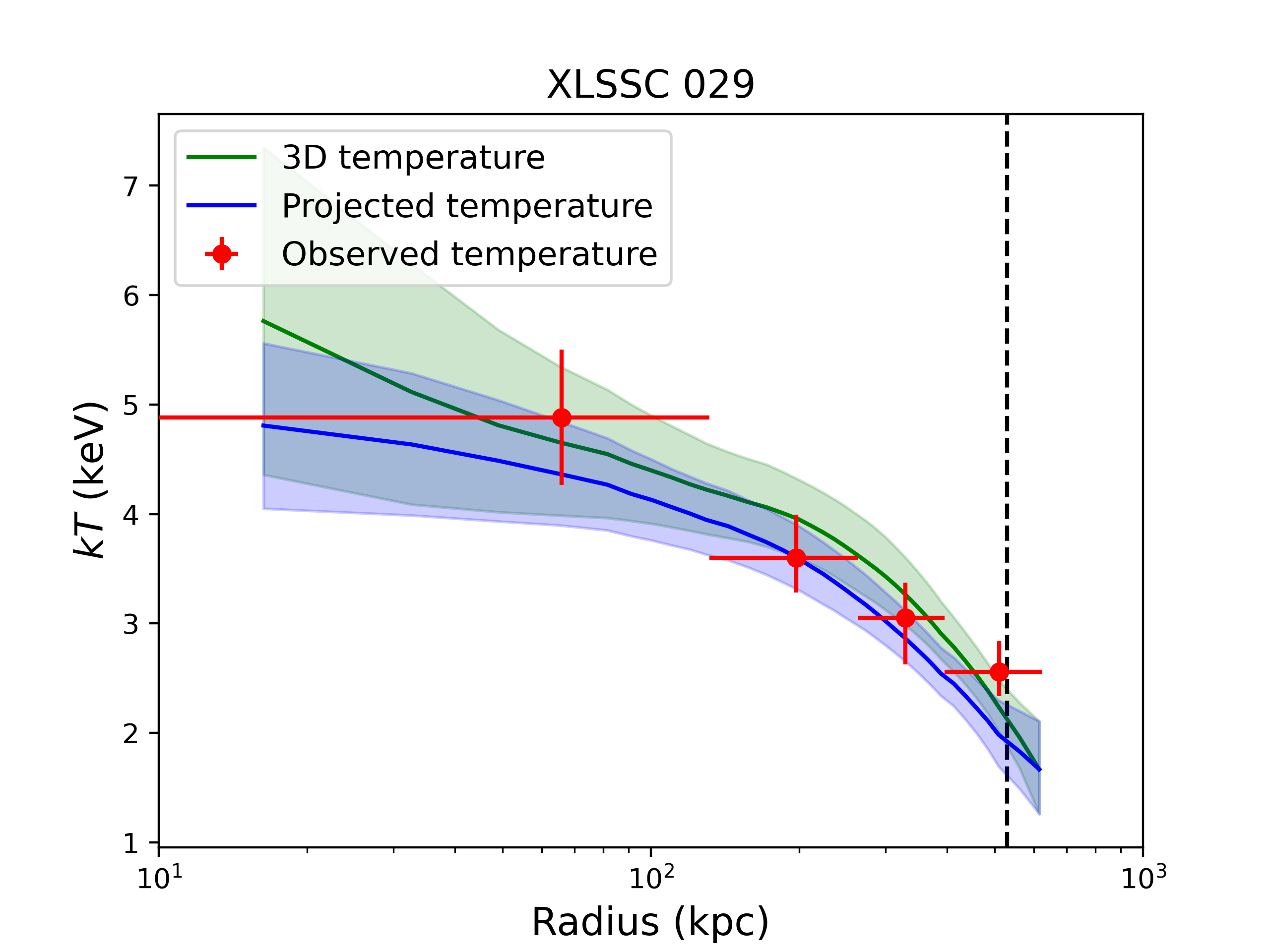}} &
	\subfloat{\includegraphics[height=2.0in]{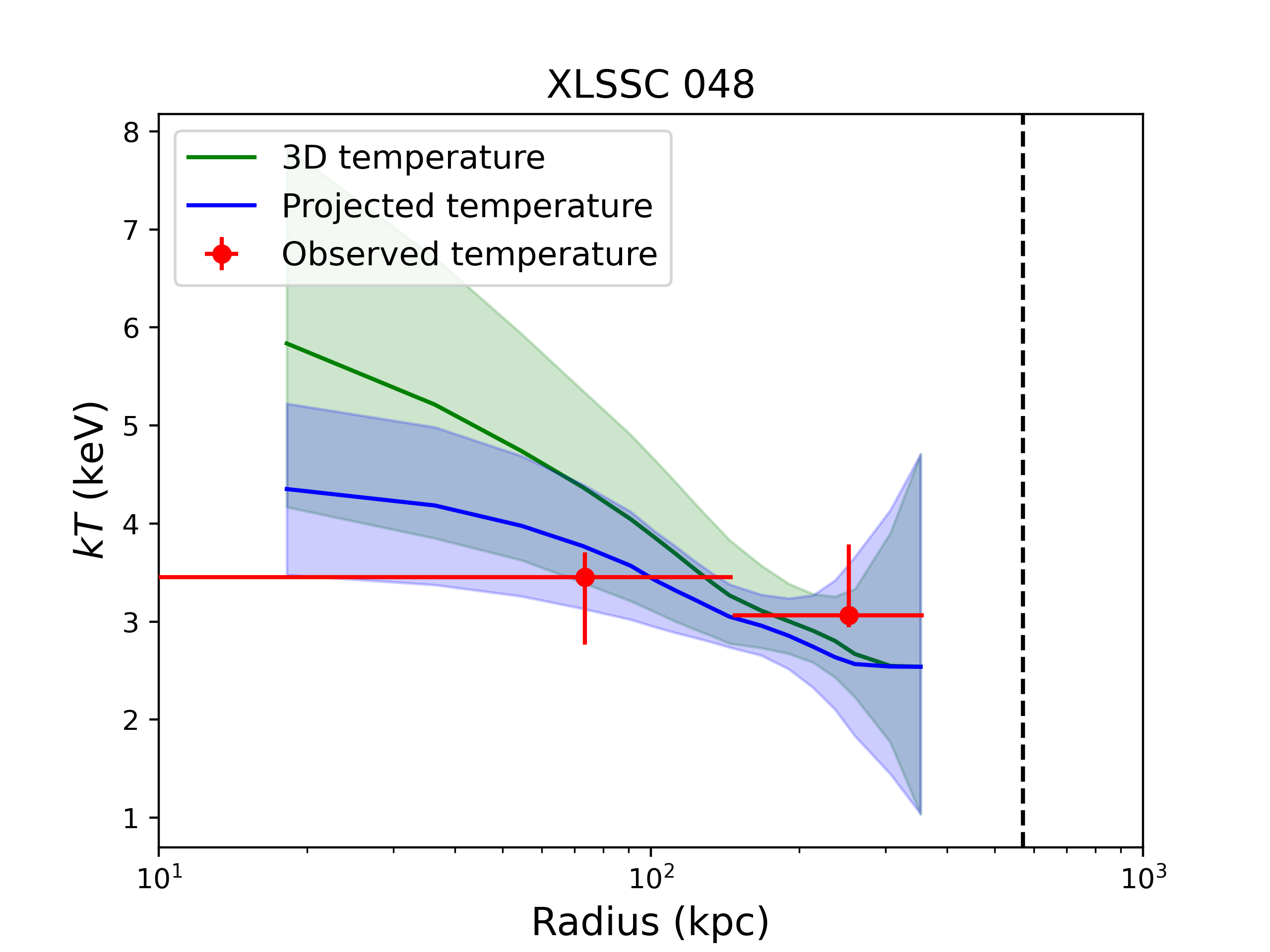}} \\
	\subfloat{\includegraphics[height=2.0in]{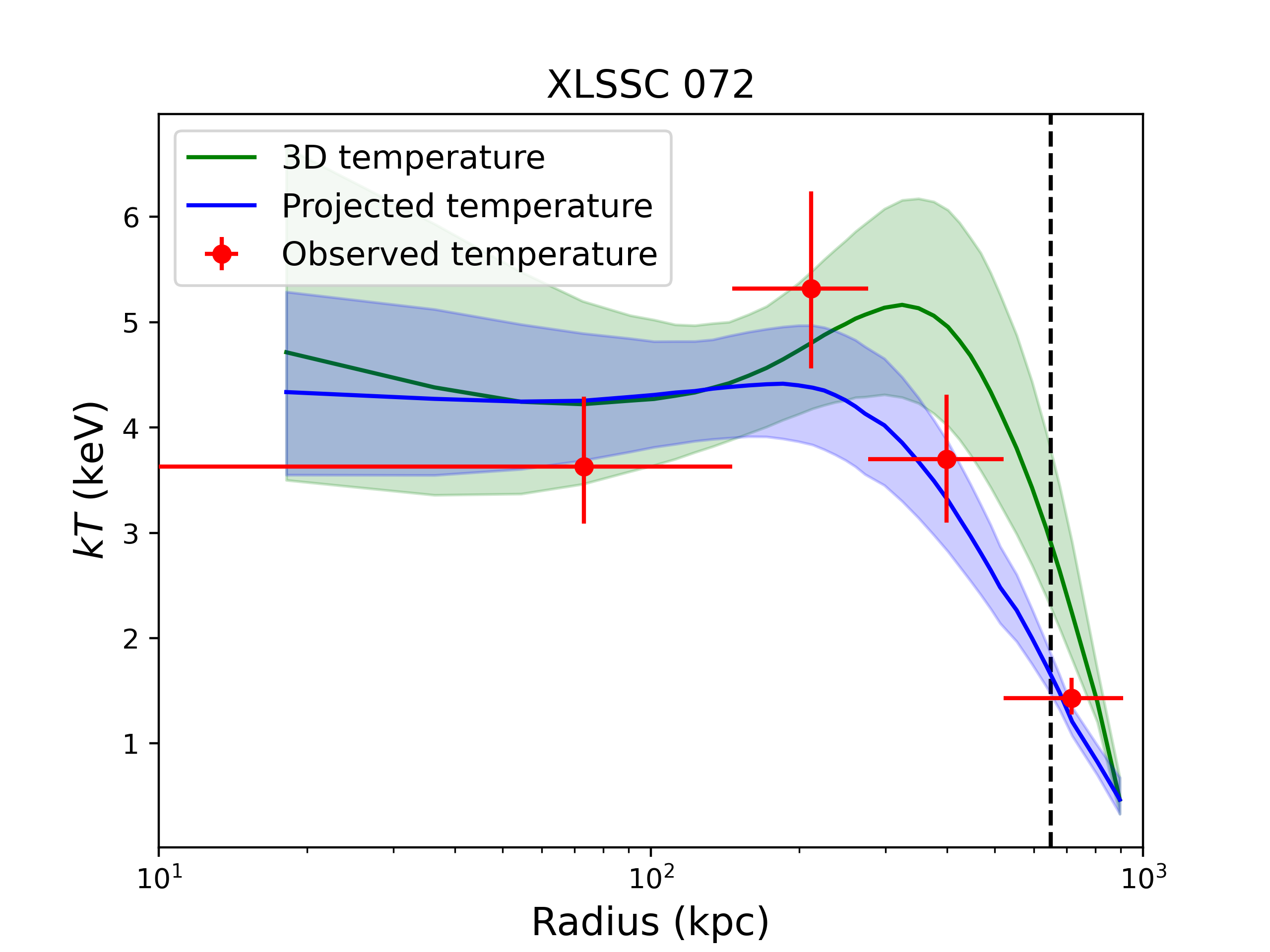}} &
	\subfloat{\includegraphics[height=2.0in]{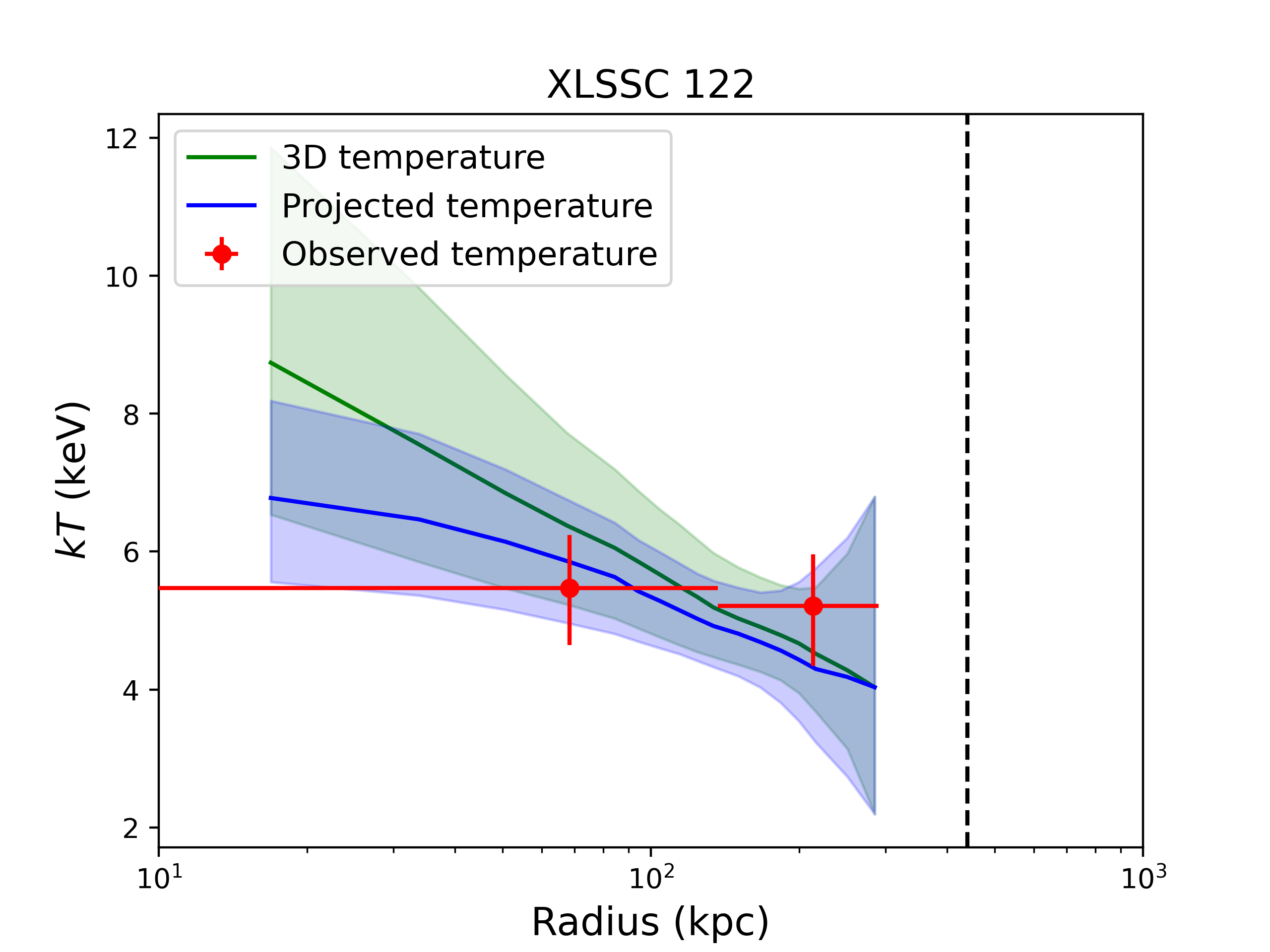}}
	\\
	\subfloat{\includegraphics[height=2.0in]{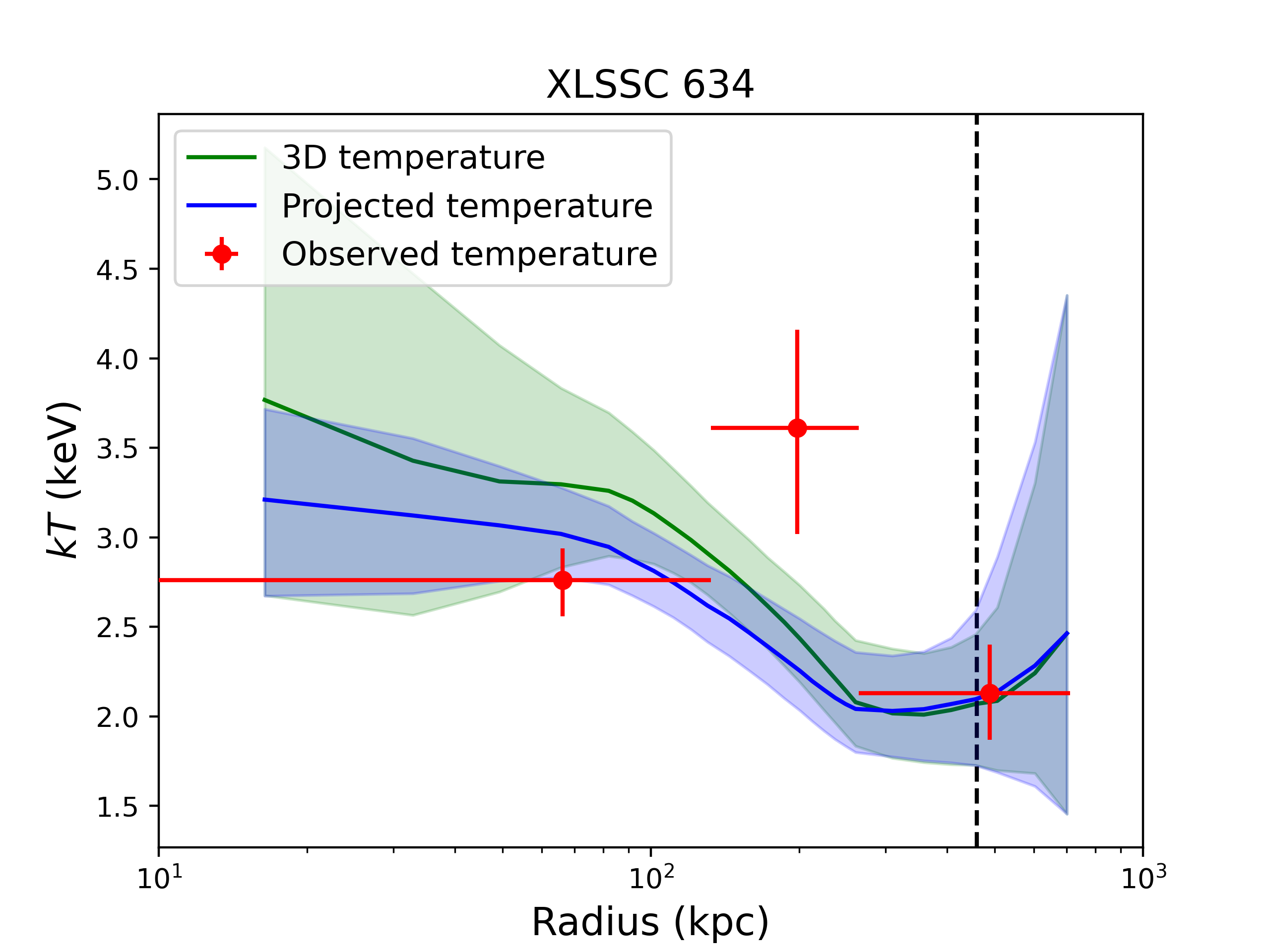}} &
	\subfloat{\includegraphics[height=2.0in]{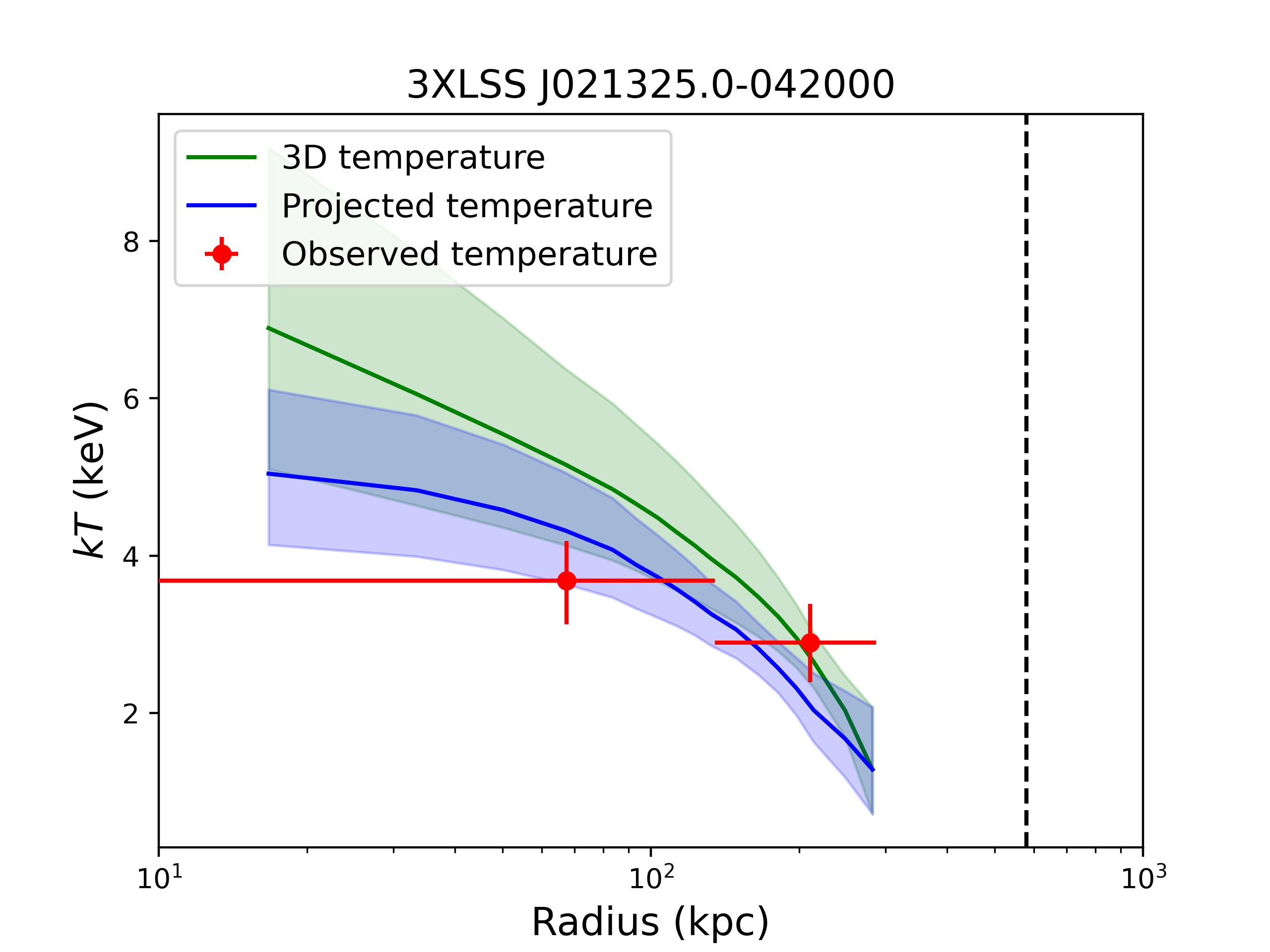}}\\
	\\
	\end{tabular}
	\caption{Temperature profiles for each of the clusters in the $z>1$ bright sample. The red points indicate the observed temperature bins for each cluster, the green lines show the 3D temperature profiles from the inversion of the hydrostatic mass equation and the blue lines show the best fitting projected temperature profile for each cluster. The dashed vertical line indicates $r_{500}$ for each cluster.}
	\label{fig:tempprofs}
\end{figure*}

The surface brightness profiles and their fit to the PSF-convolved model are shown in Figure \ref{fig:EMprofs} and the observed temperature profile, along with the best fitting projected and deprojected models are shown in Figure \ref{fig:tempprofs}.

\subsection{Calculating masses}
The cluster's mass is calculated from Equation \ref{eqn:nfw} using the chains of $r_{s}$ and $c$.

The gas mass is obtained by integrating the model gas density profiles:
\begin{align}
M_{\rm gas}(<r) = \int^{r}_{0} 4 \pi r^{2} \rho_{\rm gas}(r) dr
\end{align}
\noindent where $\rho_{\rm gas}=\mu m_{p} (n_{e} + n_{p})$, with $n_{e}=1.17n_{p}$ where $n_{p}$ is the number density of protons.

\section{Results \& Discussion}
\label{sec:results}
\subsection{Cluster Properties}
The measured thermodynamic properties of the ICM for each cluster are given in Table \ref{table:physprop}. Several of the clusters in the sample have had their thermodynamic properties studied previously. For XLSSC 029, our hydrostatic mass is in agreement with that measured in \citet{2008Maughan} ($1.4^{+0.4}_{-0.3}\times10^{14}M_{\odot}$ compared to $1.3^{+0.9}_{-0.3}\times10^{14}M_{\odot}$), despite differences in methodology.


The X-ray properties of XLSSC 122 have been thoroughly studied due to its high redshift of $z\sim2.0$ (\citealt[][hereafter XXL Paper \Rmnum{5}]{2014Mantz}, \citetalias{2018Mantz}). Here we compare the values we measure to the values measured for various properties in \citetalias{2018Mantz}. Temperature measurements are consistent ($5.0\pm0.7$ keV compared to $6.3^{+0.9}_{-0.7}$ keV). The mass estimates are not consistent, with the mass we find far higher ($2.2^{+3.5}_{-0.9}\times10^{14}M_{\odot}$ compared to $6.3\pm1.5\times10^{13}M_{\odot}$). The mass in \citetalias{2018Mantz} is determined by using the gas mass profile of XLSSC 122, and assuming a fiducial value for the value of the gas mass fraction. The gas mass within $r_{500}$ is also lower than we measure ($7.9\pm1.9\times10^{12}M_{\odot} $ compared to $1.8^{+0.7}_{-0.3}\times10^{13}M_{\odot}$). However, increasing the value of $r_{500}$ from $295\pm23$ kpc (the value used in \citetalias{2018Mantz}) to $440^{+170}_{-80}$ kpc as we find, brings these measurements close to consistency.

\begin{table*}
    \centering
	\begin{tabular}{l c c c c c c}
		\hline
		Cluster & z & $T_{\rm 300\,kpc}$ (keV) & $T_{r_{500}}$ (keV) & $M_{500}$ ($10^{14} M_{\odot}$) & $M_{\rm gas,500}$ ($10^{13}M_{\odot}$) & $f_{\rm gas,500}$  \\
		\hline
		XLSSC 029 & 1.05 & $4.2^{+0.5}_{-0.3}$ & $3.9^{+0.4}_{-0.3}$ & $1.4^{+0.4}_{-0.3}$ & $2.6^{+0.4}_{-0.2}$ & $0.18\pm0.02$ \\
		XLSSC 048 & 1.01 & $3.3^{+0.5}_{-0.4}$ & $2.9\pm0.3$ & $1.7^{+2.6}_{-0.8}$ & $1.2^{+0.3}_{-0.2}$ & $0.07\pm0.04$ \\
		XLSSC 072 & 1.00 & $4.9^{+0.8}_{-0.6}$ & $4.5\pm0.6$ & $2.4^{+1.7}_{-0.8}$ & $3.8^{+1.4}_{-0.8}$ & $0.16\pm0.03$ \\
		XLSSC 122 & 1.99 & $5.4^{+0.6}_{-0.8}$ & $6.3^{+0.9}_{-0.7}$ & $2.2^{+3.5}_{-0.9}$ & $1.8^{+0.7}_{-0.3}$ & $0.09\pm0.04$ \\
		XLSSC 634 & 1.08 & $3.7^{+0.6}_{-0.5}$ & $3.2^{+0.5}_{-0.4}$ & $1.0^{+0.5}_{-0.2}$ & $2.8^{+0.5}_{-0.3}$ & $0.29^{+0.06}_{-0.07}$ \\
		3XLSS J021325.0-042000 & 1.20 & $3.6\pm0.5$ & $3.6^{+0.5}_{-0.4}$ & $2.1^{+2.6}_{-0.9}$ & $2.6^{+1.1}_{-0.7}$ & $0.12\pm0.04$ \\
		\hline
	\end{tabular}
	\caption{Clusters and their physical properties. Column 1 gives the cluster name; Column 2 gives the redshift for each cluster; Column 3 gives the temperature of the ICM enclosed within 300 kpc from spectral fitting; Column 4 gives the hydrostatic mass within $r_{500}$; Column 5 gives the gas mass within $r_{500}$ and Column 6 gives the gas mass fraction within $r_{500}$.}
	\label{table:physprop}
\end{table*}

\subsection{Scaling Relations and Comparisons with SZ selected clusters}
\label{sec:scalingrelations}
In this section we look at several scaling relations for high redshift clusters, comparing the XXL clusters to two samples of SZ-selected clusters. Alongside the 6 $z>1 $ clusters, we also include the properties of XLSSC 102 which is at $z=0.97$ \citepalias{2020Ricci}. These SZ-selected samples consist of clusters at $z>0.9$ with hydrostatic mass estimates, and our sample of X-ray selected clusters is complementary due to its lower masses. We plot them alongside several scaling relations from other work, assuming self-similar evolution. We do not attempt to fit a best fit line using our own data, due to the narrow mass range and relatively weak mass constraints due to data quality and small sample size. There is also no attempt to correct the scaling relations for selection effects, given the sample size and large error bars.

\begin{figure}
	\centering
	\includegraphics[width=0.99\columnwidth]{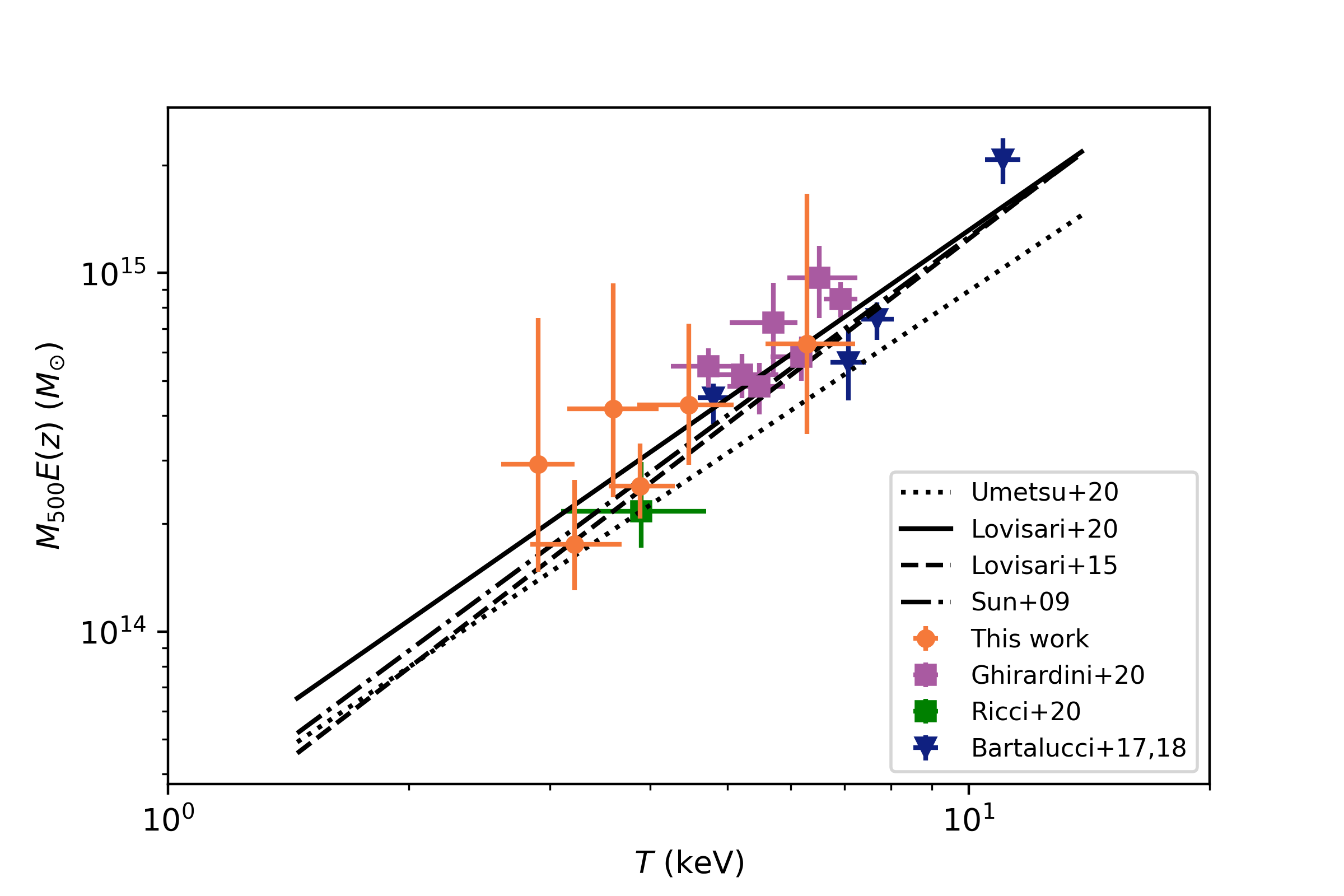}
	\caption{$M_{500}-T$ relation for our high redshift XXL clusters (orange circles) and XLSSC 102 \citepalias{2020Ricci} (green squares), alongside SZ clusters from \citet{2017Bartalucci}, \citet{2018Bartalucci} (blue triangles) and \citet{2020Ghirardini} (purple squares). Included are $M_{500}-T$ relations from \citet{2020Umetsu}, \citet{2020Lovisari}, \citet{2015Lovisari} and \citet{2009Sun}.}
	\label{fig:MT}
\end{figure}

\begin{figure}
	\centering
	\includegraphics[width=0.99\columnwidth]{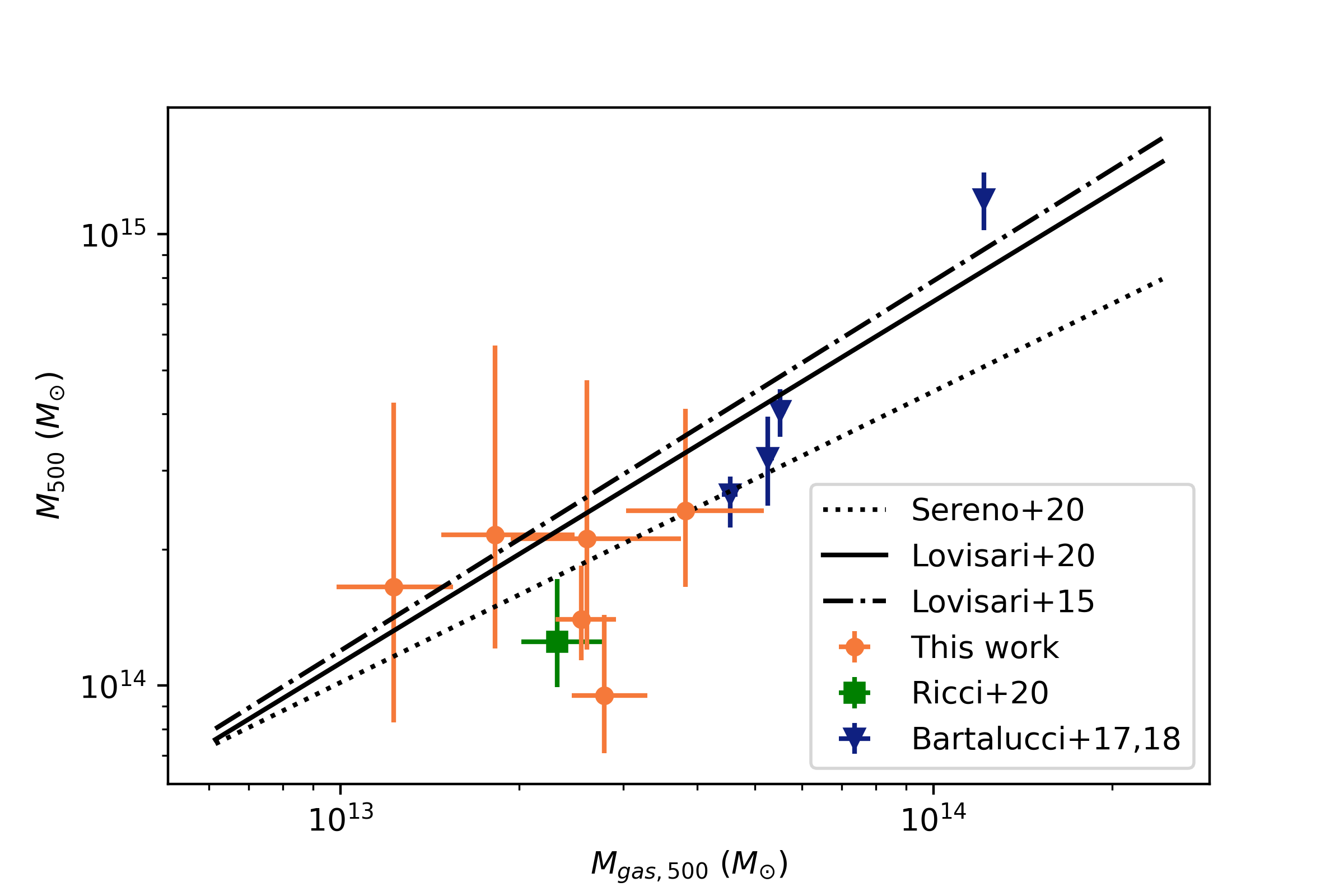}
	\caption{$M_{500}-M_{\rm gas, 500}$ relation for our high redshift XXL clusters (orange circles) and XLSSC 102 \citepalias{2020Ricci} (green squares), alongside SZ clusters from \citet{2017Bartalucci}, \citet{2018Bartalucci} (blue triangles). Included are $M_{500}-M_{\rm gas, 500}$ relations from \citet{2020Sereno}, \citet{2020Lovisari} and \citet{2015Lovisari}.}
	\label{fig:MMgas}
\end{figure}

\begin{figure}
	\centering
	\includegraphics[width=0.99\columnwidth]{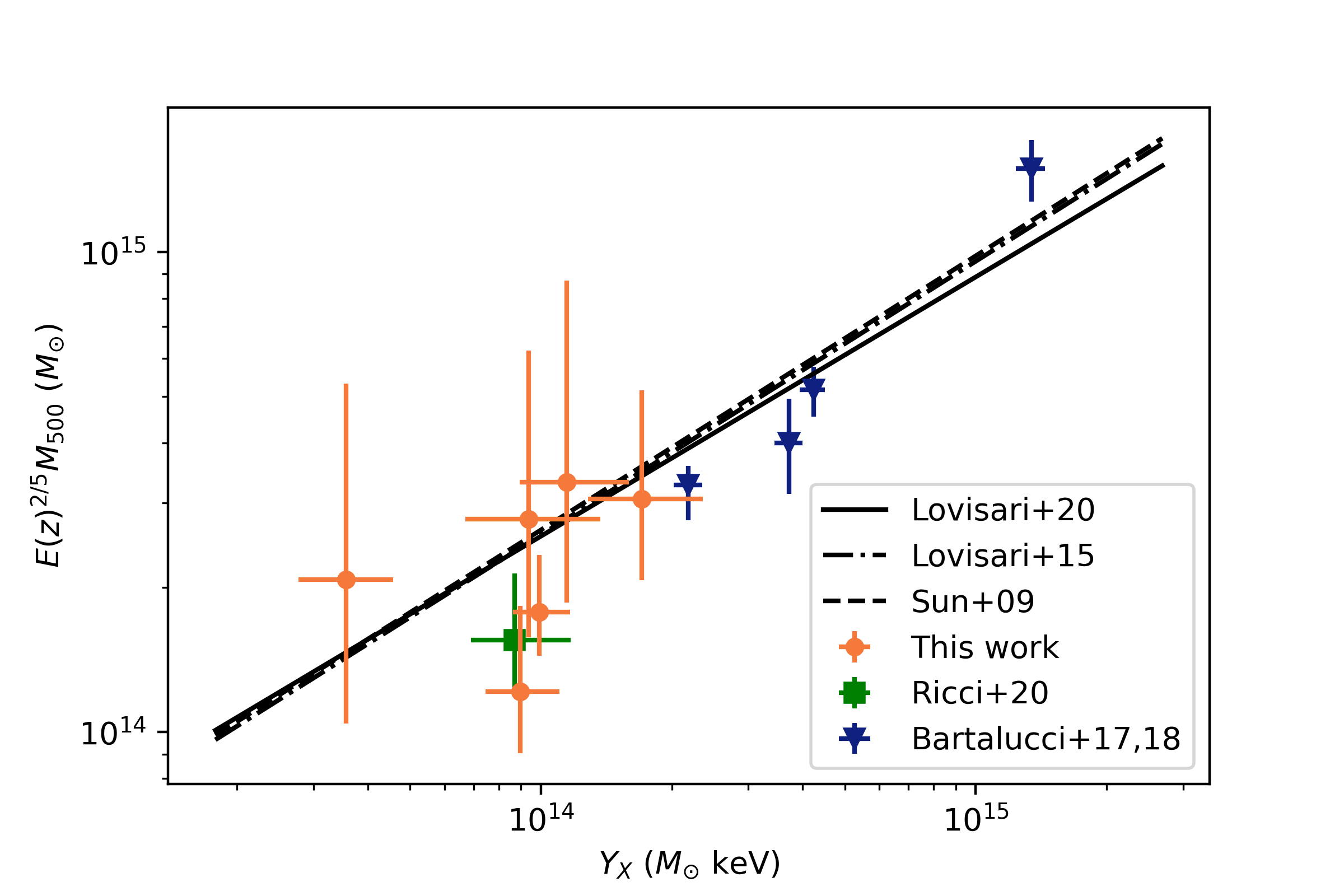}
	\caption{$M_{500}-Y_{X}$ relation for our high redshift XXL clusters (orange circles) and XLSSC 102 \citepalias{2020Ricci} (green squares), alongside SZ clusters from \citet{2017Bartalucci}, \citet{2018Bartalucci} (blue triangles). Included are $M_{500}-Y_{X}$ relations from \citet{2020Lovisari}, \citet{2015Lovisari} and \citet{2009Sun}.}
	\label{fig:MYx}
\end{figure}

As SZ signal tightly correlates with mass, and as several wide area surveys have been performed with current SZ telescopes, high redshift clusters are increasingly detected with their SZ signal rather than with X-rays. The first SZ sample we use is composed of five South Pole Telescope (SPT) SZ selected clusters with $M_{500}>5\times10^{14} M_{\odot}$ and $z\sim0.9$ from \citet{2017Bartalucci} and \citet{2018Bartalucci}. Hydrostatic masses were calculated for each of these clusters using either \emph{Chandra} or \emph{XMM-Newton} data. One of these is omitted from Figures \ref{fig:MT}, \ref{fig:MYx} and \ref{fig:MMgas}, as its mass was calculated via an extrapolation \citep{2018Bartalucci}. The second is another sample of SPT SZ selected clusters from \citet{2020Ghirardini}. Their sample consists of seven SPT-selected clusters at $z>1.2$ and with mass greater than $3\times10^{14}M_{\odot}$. Hydrostatic masses were calculated for these clusters using \emph{XMM-Newton} data.


In Figure \ref{fig:MT}, we plot the mass-temperature ($M_{500}-T$) relation for the XXL clusters, along with the two SZ samples. The figure also includes reference $M_{500}-T$ relations taken from the literature, which were derived from cluster samples defined with a variety of methods and varying mass estimation techniques. \citet{2020Umetsu} uses a subset of 105 XXL clusters which have both measured X-ray temperatures and weak lensing masses from the Hyper Suprime-Cam (HSC) Subaru Strategic Program. \citet{2020Lovisari} uses a sample of Planck SZ-selected masses with hydrostatic mass estimates. \citet{2015Lovisari} uses a sample of 20 groups, combined with additional groups and clusters from HIFLUGCS \citet{2002Reiprich} to form a larger sample of 82 with hydrostatic mass estimates. Finally, \citet{2009Sun} combine a sample of 43 groups with 14 clusters from \citet{2009Vikhlinin}, each with hydrostatic mass estimates. 

The X-ray selected clusters extend the parameter space at high redshift to slightly cooler temperatures and lower masses, as can be seen in Figure \ref{fig:MT}. The lower temperatures of X-ray clusters make sense, as the larger area and higher sensitivity to high-mass clusters of SZ surveys leads to the detection of hotter clusters. XLSSC 122, the highest redshift cluster in the sample, appears similar to the distant SZ-selected clusters in this plot. It also has an SZ detection \citepalias{2014Mantz}. The high redshift clusters show no deviation from the low-redshift scaling relations, and appear broadly consistent with the self-similar evolution of the low-redshift scaling relations.


Figure \ref{fig:MMgas} shows the $M_{500}-M_{\rm gas, 500}$ relation for the $z>1$ bright XXL clusters and the \citet{2018Bartalucci} clusters. Here, we include the scaling relation from \citet{2020Sereno}, who use the same cluster sample with weak lensing masses as \citet{2020Umetsu}. Each of these scaling relations are found to be shallower than would be expected from self-similarity. In general, the data are in good agreement with the plotted scaling relations. XLSSC 634 lies farthest from the plotted scaling relations, and has an unusually high gas mass for its hydrostatic mass.

$M_{500}$ and $M_{\rm gas, 500}$ can be used to derive the gas mass fraction $f_{\rm gas, 500}$. The average across the sample is $0.14^{+0.04}_{-0.07}$. This is consistent with the gas mass fraction found for the weak-lensing calibrated gas fraction of the XXL groups and clusters of $0.11\pm0.05$ \citep{2020Sereno}. 


$Y_{X}$, a measure of the total thermal energy in the ICM, is often used as a low scatter mass proxy \citep{2006Kravtsov}. Figure \ref{fig:MYx} shows the $M_{500}-Y_{X}$ relation for the clusters in this work and SZ selected clusters from \citet{2018Bartalucci}. The clusters are consistent with the plotted relations, suggesting that $M_{500}-Y_{X}$ holds as a reliable method of estimating the total masses of groups and clusters, even at high redshift.

\subsection{Dynamical state of clusters and reliability of the hydrostatic assumption}
\label{sec:dynamicalstate}
Throughout this work we have assumed that the clusters are in hydrostatic equilibrium in order to estimate their masses. The presence of non-thermal pressure sources associated with turbulent and bulk motions in the ICM can lead to biases in the calculation of the hydrostatic mass.  Cosmological hydrodynamic simulations of galaxy clusters have consistently shown that hydrostatic equilibrium masses underestimate true masses by 10-30\%, depending on the physics and thermodynamics occurring within the ICM and the aperture the mass is measured \citep{2014Rasia}. Simulations have also shown that ICM temperature inhomogeneities can be responsible for 10-15\% of the bias, and that when these unresolved structures are present, X-ray measurements are biased low because X-ray detectors have higher efficiency in the soft band \citep{2004Mazzotta}. In this section, we investigate the dynamical states of clusters in the sample ascertain whether the assumption of hydrostatic equilibrium holds.

\subsubsection{X-ray peak-BCG offset}
Due to the compact nature of the clusters relative to the \emph{XMM-Newton} PSF, to quantify the relaxation state of the cluster we measure the distance between the peak of the X-ray emission and the position of the brightest cluster galaxy (BCG) within the cluster. In a cluster, the BCG should preferentially lie at the centre due to dynamical friction. The X-ray emitting gas provides an observational tracer of the cluster potential, and its peak should align with the bottom of the potential.

To find the peak of the emission, images were lightly smoothed with a Gaussian function of radius 3 pixels. The peak then corresponds to the centre of the pixel where the smoothed counts were the highest. BCG positions are taken from \citet[][hereafter XXL Paper \Rmnum{15}]{2016Lavoie} for XLSSC 029 and XLSSC 072, and from \citet{2020Willis} for XLSSC 122. The BCG positions of XLSSC 048 and 3XLSS J021325.0-042000 were measured from HSC images, while a Dark Energy Survey image was used for XLSSC 634. The results from this process are displayed in Table \ref{table:bcgxp}. The offsets are given in terms of $r_{500}$, as this enables a comparison with results from other samples, as it offers a suitable normalisation method based on the mass of the cluster.


\begin{table*}
    \centering
    \scalebox{0.88}{
	\begin{tabular}{l c c c c c c c c c}
		\hline
		Cluster & z & $r_{500}$ (kpc) & BCG RA & BCG Dec & X-ray Peak RA & X-ray Peak Dec & Offset ($\arcsec$) & Offset ($r_{500}$) & $w$ ($10^{-3} r_{500}$) \\
		\hline
		XLSSC 029 & 1.05 & $530^{+50}_{-40}$ & 36.0174 & -4.2240 & 36.0171 & -4.2248 & 3.1 & 0.047 & $11.1\pm2.2$ \\
		XLSSC 048 & 1.01 & $570^{+210}_{-120}$ & 35.72025 & -3.47304 & 35.7250 & -3.4757 & 19.6 & 0.28 & $9.4\pm2.1$ \\
		XLSSC 072 & 1.00 & $650^{+120}_{-80}$ & 33.850 & -3.7256 & 33.8501 & -3.7252 & 1.5 & 0.018 & $29.4\pm3.3$ \\
		XLSSC 122 & 1.99 & $440^{+170}_{-80}$ & 34.43422 & -3.75880 & 34.4346 & -3.7582 & 2.6 & 0.049 & $20.2\pm3.3$ \\
		XLSSC 634 & 1.08 & $460^{+70}_{-40}$ & 355.69130 & -54.18480 & 355.6921 & -54.1843 & 3.4 & 0.060 & $7.6\pm1.9$ \\
		3XLSS J021325.0-042000 & 1.20 & $580^{+180}_{-100}$ & 33.35192 & -4.33395 & 33.3512 & -4.3342 & 2.7 & 0.039 & $30.7\pm3.8$ \\
		\hline
	\end{tabular}
	}
	\caption{Clusters and their BCG offsets from the peak of X-ray emission.}
	\label{table:bcgxp}
\end{table*}

\citet{2009Sanderson} and \citet{2016Rossetti} define a cluster as dynamically relaxed if the offset is between the BCG and X-ray peaks is $<0.02r_{500}$ and likely disturbed if the offset is $>0.02r_{500}$. Their work is mostly done with \emph{Chandra} observations and at far lower redshifts. We are limited not only by the extent of the \emph{XMM} PSF but, also the high redshifts of the sample. Instead, we will consider clusters with offsets greater than $0.05r_{500}$ as unrelaxed, and clusters with lower offsets relaxed as is done in \citetalias{2016Lavoie} to account for these factors. However, due to the redshifts of clusters in the sample, this offset is not a robust measurement, as the scales we are probing are at the limit of the resolution of the \emph{XMM} detectors. With this classification, XLSSC 072 and 3XLSS J021325.0-042000 would be considered relaxed, although this method of measuring the dynamical state tells us nothing about line of sight mergers. On the other hand, XLSSC 048 appears to be extremely unrelaxed. XLSSC 634 also appears to be unrelaxed, whilst XLSSC 029 and XLSSC 122 are marginal cases. XLSSC 122 is shown to be disturbed in \citetalias{2018Mantz}, who used the difference between the X-ray and SZ peaks. XLSSC 048 is the most unrelaxed cluster in the sample.

Based on this diagnostic, which is crude for high redshift clusters, most of the objects in the subsample appear unrelaxed. This is consistent with the evidence for clusters being less relaxed at high redshift, and underlines the challenge in getting precise and accurate hydrostatic masses for these clusters. However, it is especially difficult to accurately quantify the relaxation state of the clusters via this method in this particular sample, as we are limited not only by the size of the \emph{XMM} PSF ($\sim6\arcsec$), but also the physical pixel sizes of the detectors ($\sim4.1\arcsec$ for the pn detector). The offsets we measure are small compared to both of these.

\subsubsection{X-ray morphological parameters}
There are a number of other useful metrics for measuring the dynamical state of galaxy clusters \citep[e.g.][]{2017Lovisari}. \citet{2017Lovisari} suggests the best for determining whether systems are dynamically relaxed or disturbed are the concentration $c_{SB}$, and the centroid shift $w$. The concentration can be measured by integrating the PSF-corrected surface brightness profile between between two different radii and the centre.

Our data is somewhat limited in the inner regions of the clusters due to their high redshift, and our fits to the profiles do not have a functional form. As a result, it is difficult to obtain a reliable measurement of the concentration, and so we do not include this here.

To measure the centroid shift, we utilise the method of \citet{2006Poole}, where the centroid shift is defined as the standard deviation of the distance between the X-ray peak and centroid. This is measured within a series of circular apertures centred on the X-ray peak starting within a radius of $r_{500}$, and decreasing in 5\% steps until $0.05r_{500}$. The errors on the centroid shift were computed utilising 100 Monte Carlo randomisations of the input image under a Poisson distribution. The values for the centroid shift are given in Table \ref{table:bcgxp}.


The value of centroid shift which determines relaxed and disturbed clusters is subjective, and varies between studies. Using the thresholds from the SZ-selected sample from \citet{2017Lovisari}: XLSSC 029, XLSSC 048 and XLSSC 634 are considered relaxed clusters, while XLSSC 072 and 3XLSS J021325.0-042000 would be considered disturbed and XLSSC 122 would have a mixed classification.

A comparison can also be made to the thresholds of lower-redshift samples of X-ray selected clusters. When compared to the REXCESS thresholds \citep{2009Pratt}: XLSSC 072, XLSSC 122 and 3XLSS J021325.0-042000 would be considered disturbed, XLSSC 634 would be considered relaxed and XLSSC 029 and XLSSC 048 are boundary cases. When compared to the boundaries used in \citet{2012Maughan}, which is a study defined by available archival \emph{Chandra} data, all clusters in our sample are considered dynamically disturbed.

There is no agreement between the classifications of the dynamical states using the BCG offset and the centroid shift methods. The values for the centroid shift are consistent with the impression given by the X-ray images in Figure \ref{fig:clusterimg}, where XLSSC 029, XLSS 048 and XLSS 634 appear the more relaxed. The fact the BCG offset classification differs so much highlights that a definitive dynamical classification is hard with the data available for this sample of clusters.

\section{Summary}
\label{sec:conclusions}
In conclusion, we investigated the AGN contamination of clusters observed after the publication of \citetalias{2018Logan}. We find 3XLSS J231626.8-533822 to be a point source misclassified as a cluster and 3XLSS J232737.3-541618 cluster to be a genuine cluster. We expect C2 cluster candidates identified from the pipeline (used in \citetalias{2018Adami}) to be $\sim$50\% clean (\citealt{2006Pierre}, \citetalias{2018Logan}). As such, the XXL pipeline is working within expectations, even at high redshift, and our conclusions from \citetalias{2018Logan} are unaltered.

We have also measured the thermodynamic properties including the hydrostatic masses in a bright subsample of X-ray selected XXL Survey galaxy clusters at $z>1$. This sample is unique as it has been investigated for AGN contamination, and pushes to lower mass at higher redshift than is usually possible. Many of our observations were heavily affected by flaring, and so we use the backward fitting method of mass estimation, assuming an NFW mass model with few free parameters in order to constrain the masses.

We investigated the $M_{500}-T$, $M_{500}-M_{\rm gas, 500}$, and $M_{500}-Y_{X}$ relations using a combination of the results from the $z>1$ bright XXL clusters, and two samples of SPT SZ selected clusters from \citet{2018Bartalucci} and \citet{2020Ghirardini}, both of which have $z>0.9$. We find that these high redshift clusters are broadly consistent with the self-similar evolution of scaling relations determined for samples of low redshift clusters.

From the metrics considered here (BCG offsets and centroid shift), it is difficult to classify the dynamical state of these clusters with any certainty. In the future, missions such as \emph{ATHENA} with deeper data will be capable of further expanding the parameter space for scaling relations at high-redshift through the detection of lower mass clusters and groups at these redshifts \citep{2020Zhang} and accurate measurement of their thermodynamic properties within $r_{500}$ with exposures of $\sim100$ ks \citep{2018Cucchetti}. However, \emph{ATHENA} lacks the spatial resolution necessary to resolve out temperature substructures and improve mass modelling for objects at high redshift, which would be delivered by a mission such as \emph{Lynx} \citep{2018LynxTeam} or AXIS \citep{2018Mushotzky}.



\section*{Acknowledgements}
XXL is an international project based around an XMM Very Large Programme surveying two 25 deg$^{2}$ extragalactic fields at a depth of $\sim6\times 10^{-15}$ erg cm$^{-2}$ s$^{-1}$ in the [0.5-2] keV band for point-like sources. The XXL website is \url{http://irfu.cea.fr/xxl}. Multi-band information and spectroscopic follow-up of the X-ray sources are obtained through a number of survey programmes, summarised at \url{http://xxlmultiwave.pbworks.com/}. The scientific results reported in this article are based in part on observations made by the \emph{Chandra} X-ray Observatory. RTD and BJM acknowledge support from Science and Technologies Facility Council (STFC) grant ST/R000700/1. CL acknowledges support from the UK STFC, along with an ESA Research Fellowship. MP acknowledges long-term support from the Centre National d'Etudes Spatiales (CNES). We thank L. Chiappetti for the technical report.

\section*{Data Availability}
The X-ray observations analysed in this work are publicly available at either the \emph{XMM-Newton} Science Archive\footnote{\url{http://nxsa.esac.esa.int/nxsa-web/\#search}}, or the \emph{Chandra} Data Archive\footnote{\url{https://cda.harvard.edu/chaser/}}. Data products derived from these are available from the author upon request.

\bibliography{bib}

\end{document}